\documentclass[showpacs, amsmath, amsfonts, amssymb, aps, superscriptaddress, nofootinbib]{revtex4}

\usepackage{pdflscape}
\usepackage{rotating}
\usepackage{tabularx}
\usepackage[dvipsnames]{xcolor}
\usepackage{comment}
\usepackage{enumitem}
\usepackage{graphicx}
\usepackage{float}
\usepackage{dcolumn}
\usepackage{bm}
\usepackage{mathrsfs}
\usepackage{amsmath}

\usepackage{amsfonts}
\usepackage{dsfont}

\usepackage{csquotes} 
\usepackage[normalem]{ulem}
\usepackage{booktabs}
\usepackage{amstext} 
\usepackage{array}   
\newcolumntype{C}{>{$}c<{$}} 
\newcolumntype{L}{>{$}l<{$}} 

\usepackage{url,hyperref}
\hypersetup{colorlinks=true,
linkcolor=blue,
citecolor=violet,
filecolor=violet,
urlcolor=blue}

\begin{document}

\title{Quasinormal-mode analysis of a massive scalar field in a Schwarzschild background via the spectral method}

\author{Davide Batic}
\email{davide.batic@ku.ac.ae}
\affiliation{Mathematics Department, Khalifa University of Science and Technology, PO Box 127788, Abu Dhabi, United Arab Emirates}

\author{Anna Chrysostomou}
\email{chrysostomou@lpthe.jussieu.fr}
\affiliation{Laboratoire de Physique Th\'eorique et Hautes \'Energies - LPTHE, Sorbonne Universit\'e, CNRS, 4 Place Jussieu, 75005 Paris, France}

\author{Alan S. Cornell}
\email{acornell@uj.ac.za}
\affiliation{Department of Physics, University of Johannesburg, PO Box 524, Auckland Park 2006, South Africa}

\author{Denys Dutykh}
\email{denys.dutykh@ku.ac.ae}
\affiliation{Mathematics Department, Khalifa University of Science and Technology, PO Box 127788, Abu Dhabi, United Arab Emirates}

\date{\today}

\begin{abstract}
\textcolor{black}{
We study the massive scalar quasinormal-mode spectral problem on a
Schwarzschild background using an extended Chebyshev spectral method. The massive dispersion relation has a two-sheeted analytic structure, which we uniformise before discretisation. This converts the radial problem into a seventh-degree polynomial eigenvalue problem while incorporating the quasinormal-mode boundary conditions at the horizon and spatial infinity. The method yields stable spectra for low multipoles, high overtones, and frequencies on both sides of the nominal mass threshold, including regimes in which WKB and continued-fraction approaches become difficult to apply. For complex frequencies, the far-field behaviour cannot be classified from the real part of the frequency alone, but depends on the complex wavenumber and its Riemann sheet. We also identify stable families of purely imaginary roots with approximately uniform spacing in their damping rates, whose physical interpretation remains open. The Schwarzschild scalar problem is intended as a controlled benchmark for massive-field quasinormal-mode calculations rather than as a direct model of tensorial Kerr ringdown. 
}
\end{abstract}

\maketitle
\section{Introduction \label{sec:intro}} 

As a perturbed black hole of mass $M$ relaxes towards equilibrium, its intermediate ringdown phase is governed by characteristic quasinormal modes (QNMs). As demonstrated in the seminal works of Vishveshwara \cite{Vishveshwara1970_scattering,Vishveshwara1970_stability}, Press \cite{Press1971}, and Leaver \cite{Leaver1985PRSLA}, these modes define an intrinsically dissipative but well-posed boundary-value problem, with purely ingoing behaviour at the event horizon and purely outgoing behaviour at spatial infinity. The corresponding eigenvalues, the discrete spectrum of quasinormal frequencies (QNFs), are denoted by 
\begin{equation}
 \Omega=\Omega_{\rm R}+i\Omega_{\rm I} \;,   \qquad  \Omega=M\omega \;.
\end{equation}
\noindent \textcolor{black}{Throughout this work, we use the time dependence $e^{-i\omega t}$. Accordingly, $\Omega_{\rm R}\equiv\Re {\rm e} \{\Omega\}$ is the oscillation frequency and $\Omega_{\rm I}\equiv\Im {\rm m} \{\Omega\}<0$ gives temporal decay, whereas $\Omega_{\rm I}>0$ would signal growth.} For static, spherically symmetric black hole spacetimes, the QNFs of a field of spin $s$ depend on the multipole number $\ell$. For each $\ell$, they form a countably infinite sequence labelled by the overtone number $n$. The eikonal \textcolor{black}{(geometric-optics)} limit corresponds to $\ell\to\infty$, whereas the highly damped regime corresponds to $n\to\infty$.

\textcolor{black}{We stress at the outset that the minimally coupled test scalar considered here is not the tensorial ringdown channel targeted by gravitational-wave analyses of compact-binary remnants. In General Relativity (GR), black hole spectroscopy is primarily based on gravitational perturbations of the remnant Kerr black hole.  Consequently, the frequencies computed below are not proposed as direct gravitational-wave templates unless a concrete interaction both excites the scalar degree of freedom and imprints it on the observable gravitational waveform \cite{Berti2009CQG,Konoplya2011RMP,Lagos2020_Anomalous}. The role of the present Schwarzschild scalar problem is instead methodological and foundational because it isolates, in a clean one-field setting, the two-sheeted dispersion relation, the modified asymptotic boundary condition, the mass threshold, sheet-dependent resonant and decaying asymptotics, and quasiresonant or quasibound behaviour. It therefore provides a controlled benchmark before addressing the observationally central problem of gravitational perturbations of Kerr black holes. Physically motivated extensions include coupled scalar--tensor perturbations, for which scalar modes can leak into the gravitational signal, massive vector or tensor sectors tied to a specified particle physics or beyond-GR model, and massive fields on Kerr backgrounds, where quasibound states are connected with superradiance and ultralight-boson phenomenology \cite{Dolan2007_MassiveScalarKerr,Arvanitaki2010PRD,BritoCardosoPani2020_Superradiance,Dolan2020_MassKerr}. Electromagnetic perturbations concern a distinct observational channel, and a connection to black hole imaging or accretion signatures would additionally require a model of the emitting plasma and its coupling to the spacetime, which is outside the vacuum test-field calculation performed here.}

The QNMs of massless scalar fields propagating on a Schwarzschild background have been explored extensively (see Refs.~\cite{Berti2009CQG,Konoplya2011RMP}, and the references therein). Introducing a nonzero field mass $\mu=mM$ changes the asymptotic value of the effective potential from zero to $\mu^2$, modifies the admissible analytic continuations at infinity, and may generate an outer well supporting long-lived configurations \cite{SimoneWill1991_MassiveScalar,Konoplya2005PLB,Ohashi2004CQG,Dolan2007_MassiveScalarKerr,Lagos2020_Anomalous}. On the real-frequency axis, $k^2=\Omega^2-\mu^2$ gives the familiar threshold, i.e. $|\Omega|>\mu$ yields a real wavenumber $k$ and corresponds to a \emph{propagative mode}, whereas $|\Omega|<\mu$ yields a purely imaginary $k$ and corresponds to an \emph{evanescent mode}. For a complex resonance frequency, however, $k$ is generically complex and this binary classification is no longer exact; the spatial character must instead be determined from the full complex wavenumber and its branch. \footnote{Percival and Dolan~\cite{Dolan2020_MassKerr} extend the terminology
\emph{propagative} and \emph{evanescent} to complex frequencies according to
the sign of $\Re {\rm e}\{\Omega^2\}-\mu^2$. We do not use this convention here. 
Since $\Re {\rm e}\{\Omega^2\}=\Omega_{\rm R}^2-\Omega_{\rm I}^2$, this criterion
differs from the real-frequency threshold proxy $\delta_{\rm R}$ employed
below and does not, for generic complex $k$, by itself specify the full asymptotic spatial behaviour.}
\textcolor{black}{ To organise the numerical tables without prejudging the Riemann sheet, we introduce only the real-frequency threshold proxy, 
\begin{equation} \label{eq:deltaR}
    \delta_{\rm R}\equiv[\Re {\rm e}\{\Omega\}]^2-\mu^2 = \Omega_{\rm R}^2 -\mu^2 \;.
\end{equation}
This quantity must not be confused with $\Re {\rm e}\{\Omega^2\}-\mu^2=\Omega_{\rm R}^2-\Omega_{\rm I}^2-\mu^2,$ and neither sign gives a complete physical classification when $\Omega_{\rm I}\neq 0$. In what follows, $\delta_{\rm R}>0$ and $\delta_{\rm R}<0$ mean only \emph{above} and \emph{below} the nominal real-frequency threshold, respectively. A quasibound interpretation additionally
requires the decaying sign at infinity and the appropriate sheet. It is never
inferred from $\delta_{\rm R}$ alone.}

The field mass also changes the analytic structure of the time-dependent response. \textcolor{black}{Massive scalar perturbations are known to develop oscillatory late-time tails associated with the branch points at $\Omega=\pm\mu$ and with the corresponding branch-cut contribution to the retarded Green function \cite{KoyamaTomimatsu2001MassiveTail,QianEtAl2022MassiveTails}. The uniformisation used below treats the same square-root dispersion relation in the frequency-domain boundary condition, but it does not construct the retarded Green function, evaluate its discontinuity across a cut, or perform the inverse-frequency transform. We therefore make no claim here about tail exponents, amplitudes, or crossover times. The present calculation concerns discrete stable spectral roots. It can identify candidate long-lived resonant or quasibound configurations through small damping rates, once their Riemann sheet and spatial asymptotics are established, but such pole data must be distinguished from the branch-cut contribution that controls the asymptotic tail.}

Semi-analytic and recurrence-based methods remain indispensable benchmarks for the massive problem, but their domains of reliability are method dependent. WKB approximations are most reliable for barrier-controlled modes and deteriorate as the local maximum and the associated turning-point structure are suppressed. Leaver-type continued fraction method (CFM) approaches, by contrast, can remain highly accurate for massive fields when the asymptotic branch is fixed consistently and the recurrence is well conditioned. They have been used to follow ordinary QNMs towards the quasiresonant regime \cite{Konoplya2005PLB,Ohashi2004CQG,AlvesPonquioMedeiros2025PRD}. \textcolor{black}{Their limitation is therefore not a generic failure at nonzero mass. The calculation becomes delicate near the threshold $k=0$, where the Coulomb exponent and the recurrence coefficients contain inverse powers of $k$, and whenever outgoing and spatially decaying analytic continuations must be tracked separately. The uniformised spectral formulation is complementary rather than a replacement. More precisely, it treats both sheets within one polynomial problem, while a branch-resolved continued fraction supplies an independent benchmark wherever it converges reliably.}

Recently, (pseudo)spectral methods have gained traction for their exponential convergence and their ability to resolve modes in parameter regions where barrier-based approximations lose accuracy and recurrence calculations require careful branch tracking. The perturbation equation is discretised in a global spectral basis, so derivatives and coefficient functions become matrices acting on the spectral coefficients. For a massless scalar field on a Schwarzschild background this gives a quadratic matrix eigenvalue problem. \textcolor{black}{The massive problem considered below retains the same numerical philosophy but requires a uniformising parameter to accommodate its two-sheeted dispersion relation. A controllable basis size $N$ permits the systematic extraction of fundamental modes and higher overtones together with refinement tests for nontrivial asymptotic structures. Although computationally more demanding than standard low-mode calculations, spectral approaches have proved useful in problems containing overdamped branches or otherwise nonstandard asymptotics; see, for example, Refs.~\cite{Jansen:2017_Overdamped,Fortuna:2020obg,DiasSantos2020_RNdSinstability,Konoplya:2022zav,Batic2024CQG,Batic2024EPJC,Batic2024PRD,Batic2025CQG,Batic2025EPJC,Batic2025PRSA,Batic2026EPJC,BaticDutykhSukaiti2026KSQC}.}

Several variants of (pseudo)spectral methods have been proposed for QNM calculations, including Chebyshev and Bernstein expansions, collocation schemes, and global polynomial discretisations. In this work, we adopt the Chebyshev-based approach developed in Refs.~\cite{Batic2024CQG,Batic2024PRD,Batic2025EPJC,Batic2025CQG,Batic2025PRSA,Batic2026EPJC}, which provides a flexible and robust framework for the massless problem. A nonzero scalar-field mass $\mu$, however, modifies both the analytic and numerical structure of the eigenvalue problem. In the standard massless formulation, factoring out the asymptotic behaviour and discretising the radial equation yields a quadratic matrix eigenvalue problem of the form
\begin{equation}\label{eq:oldpencil}
\mathcal{P}_2(\Omega)\,\mathbf{u}\equiv \left(\Omega^2 M_2 + \Omega M_1 + M_0\right)\mathbf{u} = 0.
\end{equation}
For the massive field, the asymptotic wavenumber becomes $k(\Omega) = \sqrt{\Omega^2-\mu^2},$ and hence, the frequency dependence is intrinsically multivalued in the complex plane. To treat this structure systematically, we introduce a uniformisation via the {\textcolor{black}{Joukowski transformation (also transliterated as Zhukovsky)}} \cite{Joukowsky1910ZFM}. This removes the square root and rewrites the problem in terms of the single complex parameter $\Lambda$. Although the original formulation is quadratic in $\Omega$, the rational dependence generated by the map contains positive and negative powers of $\Lambda$. After the denominators are cleared, the discretised problem becomes a seventh-degree polynomial eigenvalue problem. The increased algebraic order therefore reflects the uniformisation of the massive dispersion relation rather than an additional physical degree of freedom.

From this analytic perspective, the wavenumber $k(\Omega)$ has branch points at $\Omega=\pm\mu$ and lives on a two-sheeted Riemann surface. The Joukowski transformation resolves this structure by encoding the two sheets in reciprocal values of the uniformising parameter and mapping the branch points to the corresponding fixed points. \textcolor{black}{Uniformisation removes the square root from the matrix polynomial and makes the sheet information algebraic. It does not, by itself, select one physical sector, because the polynomial problem may return roots from either sheet. A physical asymptotic classification therefore requires retaining $\Lambda$ together with the mapped frequency $\Omega$.}
 
Our approach is novel in two respects. First, it extends the spectral method from a quadratic to a seventh-degree polynomial eigenvalue problem, and second, it provides a unified framework in which the complex-plane geometry, Riemann-surface structure, and physical interpretation of the spectrum are treated consistently. To distinguish this formulation from earlier implementations, we refer to it as the \emph{extended spectral method}.

The paper is organised as follows: Section~\ref{sec:context} formulates the massive scalar eigenvalue problem on a Schwarzschild background. Section~\ref{sec:method} constructs the uniformised polynomial problem and specifies the requirements for a sheet-resolved asymptotic classification. Section~\ref{sec:results} presents the numerical results and benchmarks, and discusses their interpretation and limitations. Section~\ref{sec:conc} summarises the conclusions and outlook.

\section{Metric and equations of motion \label{sec:context}}

\begin{figure}[t]
    \centering
    \includegraphics[width=0.7\linewidth]{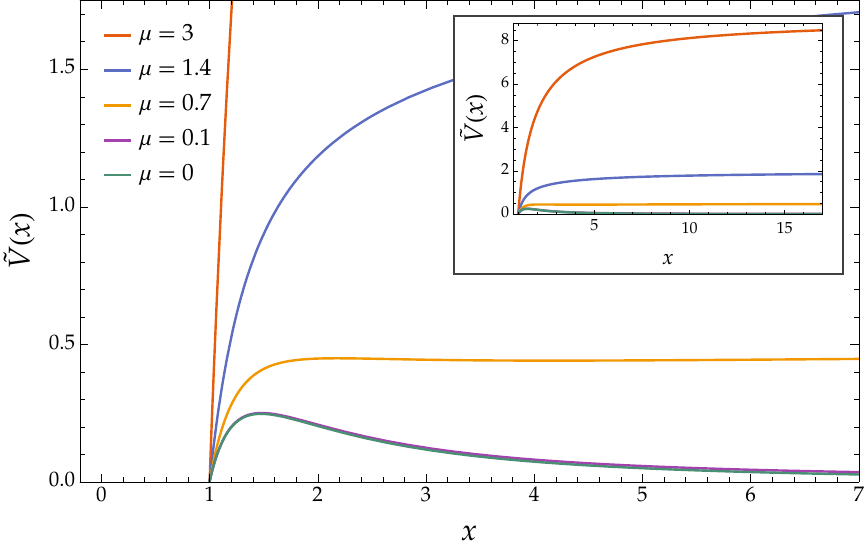}
    \caption{\textit{Effective potential defined in Eq.~\eqref{eq:VxpotM2}, with $\widetilde V(x)\to\mu^2$ as $x\to\infty$. In the massless case, the barrier peak associated with the photon sphere is located at $x=3/2$. For $\ell=2$, the curve at $\mu=0.7$ still possesses a {\textcolor{black}{shallow local maximum--minimum pair because $\mu^2<\mu_c^2$. Both extrema lie below the asymptotic level $\widetilde V(\infty)=0.49$, so the curve no longer has a conventional barrier rising above the continuum threshold. The two extrema coalesce at $\mu_c\simeq 0.7401$ and are absent for $\mu>\mu_c$.}}}}
    \label{fig:Vplot}
\end{figure}

\begin{figure}[t]
    \centering
    \includegraphics[width=0.4\linewidth]{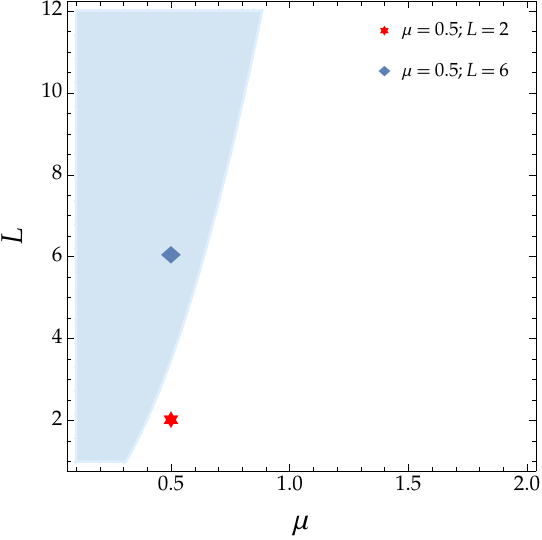} \hspace{1.15cm}
\includegraphics[width=0.45\linewidth]{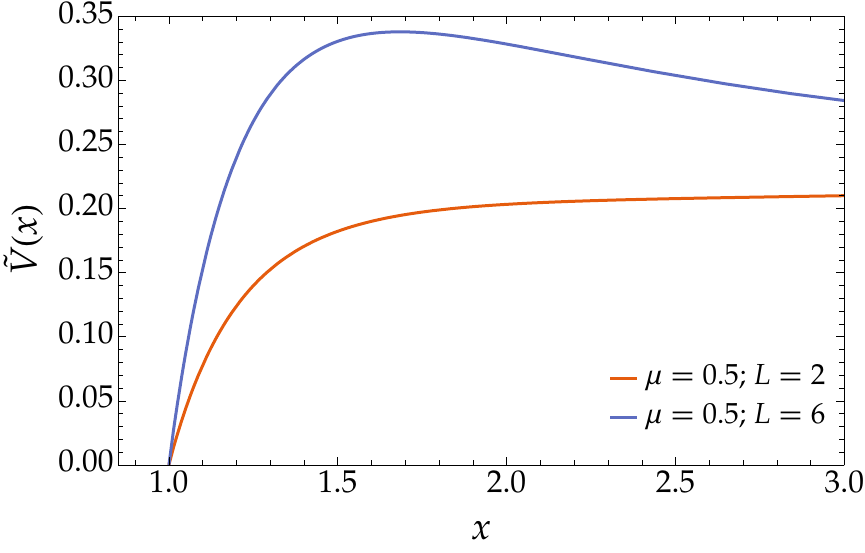}
    \caption{\textit{Left: Region of the $(\mu,L)$--plane in which the effective potential in Eq.~\eqref{eq:VxpotM2} develops a local maximum followed by a local minimum for $x>1$. The shaded area satisfies Eq.~\eqref{muineq}. The red star and blue diamond mark examples with $\mu^2>\mu_c^2(L=2)$ and $\mu^2<\mu_c^2(L=6)$, respectively, where $L=\ell(\ell+1)$. Right: Effective potentials for the two marked points, corresponding to $\ell=1$ and $\ell=2$. Only the $\ell=2$ case has the maximum--minimum trapping structure.}}
    \label{fig:Ineq}
\end{figure}

To formulate the QNM eigenvalue problem and fix our conventions, we briefly review the equation governing a massive scalar field on the Schwarzschild background. In Planck units, where $c=G_N=\hbar=1$, the Klein--Gordon equation for the field {\textcolor{black}{$\Psi=\Psi(t,r,\vartheta,\varphi)$}} in the Schwarzschild metric is
\begin{equation}
\frac{1}{\sqrt{-g}}\partial_\nu\!\left(g^{\mu\nu}\sqrt{-g}\,\partial_\mu\Psi\right)-m^2\Psi=0 \;,
\end{equation}
where the corresponding line element is
\begin{equation}
ds^2=-f(r)dt^2+\frac{dr^2}{f(r)}+r^2(d\vartheta^2+\sin^2{\vartheta} d\varphi^2)\; ,\quad
f(r)=1-\frac{2M}{r} \;.
\end{equation}
\textcolor{black}{We separate the field according to}
\begin{equation}\label{eq:separation}
\textcolor{black}{\Psi(t,r,\vartheta,\varphi)=e^{-i\omega t}Y_{\ell m}(\vartheta,\varphi)\frac{\psi(r)}{r} \;,\quad\ell\geq 0 \;,\quad |m|\leq\ell} \;,
\end{equation}
\textcolor{black}{where $Y_{\ell m}(\vartheta,\varphi)$ denotes the spherical harmonics.  The sign convention in Eq.~\eqref{eq:separation} is used throughout the manuscript.} The corresponding radial equation takes the form \cite{Konoplya2005PLB}
\begin{equation}\label{radial}
f(r)\frac{d}{dr}\left(f(r)\frac{d\psi}{dr}\right)+\left[\omega^2-V(r)\right]\psi(r)=0, \end{equation}
with the effective potential
\begin{equation} \label{eq:Vpot}
V(r)=f(r)\left(\frac{L}{r^2}+\frac{2M}{r^3}+m^2\right),\quad L=\ell(\ell+1) .
\end{equation}
In what follows, it is convenient to introduce the rescaling $x = r/(2M)$. We also make use of the dimensionless quantities $\Omega=M\omega$ and $\mu=Mm$. This allows us to rewrite Eq.~\eqref{eq:Vpot} as
\begin{equation} \label{eq:Vxpot}
V(x)=\left(1 - \frac{1}{x} \right) \left[\frac{L}{4 M^2 x^2}+\frac{1}{4 M^2 x^3}+\left( \frac{\mu}{M}\right)^2\right].
\end{equation}
We then define $\widetilde{V}(x)\equiv M^2V(x)$ to obtain the potential in fully dimensionless form,
\begin{equation} \label{eq:VxpotM2}
\widetilde{V}(x) = \left(1 - \frac{1}{x} \right) \left(\frac{L}{4 x^2}+\frac{1}{4 x^3}+ \mu^2\right).
\end{equation}
Fig.~\ref{fig:Vplot} shows this potential for $\ell=2$ and increasing values of $\mu$. The rescaling $x=r/(2M)$ places the event horizon at $x=1$. In the eikonal regime $\ell\gg 1$, the angular-momentum term $L/x^2$ dominates, and the potential maximum is associated with the unstable circular null geodesic. By contrast, when $\mu\gg\ell$, the mass term $\mu^2$ dominates and fixes the asymptotic value of the potential. \textcolor{black}{As the mass term becomes comparable to the angular contribution, two distinct transitions must be separated. First, the local maximum may fall below the asymptotic plateau $\widetilde V(\infty)=\mu^2$, so that the potential no longer presents a conventional barrier above the continuum threshold. Second, at a larger mass, the local maximum and minimum coalesce and disappear. In the resulting mass-dominated regime, the usual photon-sphere interpretation is lost. Many low-multipole roots satisfy $\delta_{\rm R}<0$, but this statement alone does not determine whether the corresponding radial solution decays or grows at infinity. That distinction requires $k(\Lambda)$ and the Riemann sheet.}
\textcolor{black}{We can determine explicitly for which parameter values the rescaled potential, Eq.~\eqref{eq:VxpotM2}, develops a local maximum followed by a local minimum for $x>1$. This stationary-point criterion is distinct from requiring the maximum to lie above the asymptotic plateau. Differentiating the potential gives}
\begin{equation}
\widetilde{V}^{'}(x)=\frac{4\mu^{2}x^{3}-2L x^{2}+3(L-1)x+4}{4x^{5}}=\frac{P(x)}{4x^{5}}.
\end{equation}
The stationary points are therefore the roots of the cubic,
\begin{equation}\label{cubic}
P(x)=4\mu^{2}x^{3}-2L x^{2}+3(L-1)x+4=0 .
\end{equation}
For $\mu>0$, the leading coefficient of $P(x)$ is positive and $P(0)=4>0$. For $0<x\leq 1$, it is easy to check that $P(x)\geq-2Lx^2+3(L-1)x+4>0$, so $P(x)$ has no roots in $(0,1]$. Since $P(x)\to-\infty$ as $x\to-\infty$ and $P(x)\to+\infty$ as $x\to+\infty$, whenever $P(x)$ has three real roots, denoted by $x_0$, $x_M$, and $x_m$, they are ordered as $x_0<0<1<x_M<x_m$. The roots $x_M$ and $x_m$ correspond, respectively, to a local maximum and a local minimum of the effective potential. Thus, the potential has this maximum--minimum structure for $x>1$ precisely when Eq.~\eqref{cubic} has three distinct real roots, equivalently when its discriminant is positive \cite{Bronshtein2015handbook}.  A straightforward calculation of the discriminant gives
\begin{equation}
\Delta(L,\mu)=4\left[-1728\mu^{4}-108(L-1)(L+1)^{2}\mu^{2}+L^{2}\left(9L^{2}+14L+9\right)\right].
\end{equation}
The condition $\Delta(L,\mu)>0$ is equivalent to $1728\mu^{4}+108(L-1)(L+1)^{2}\mu^2-L^{2}(9L^{2}+14L+9)<0$, which is a quadratic inequality in $\mu^2$ with two real roots. The physically relevant $\mu>0$ root is
\begin{equation}\label{root}
\mu_{c}^{2}(L)=\frac{1}{288}\left[\sqrt{3}(3L^{2}+2L+3)^{3/2}-9(L-1)(L+1)^{2}\right]>0,
\end{equation}
so that, for $L\geq 2$ corresponding to $\ell\geq 1$, the effective potential has a local maximum at $x_{M}>1$ followed by a local minimum at $x_{m}>x_{M}$ when
\begin{equation}\label{muineq}
0<\mu^{2}<\mu_{c}^{2}(L).
\end{equation}
\textcolor{black}{Let us consider an explicit example. For $\ell=2$, we have $L=6$ and $\mu_c^2\simeq 0.547753$, hence $\mu_c\simeq 0.740103$. The value $\mu=0.7$ satisfies $\mu^2=0.49<\mu_c^2$, so Eq.~\eqref{muineq} correctly predicts a local maximum at $x_M\simeq 2.1776$ followed by a local minimum at $x_m\simeq 4.1696$. Their values are $\widetilde V(x_M)\simeq 0.44914$ and $\widetilde V(x_m)\simeq 0.44069$, both below the asymptotic plateau $\widetilde V(\infty)=0.49$. Thus the maximum--minimum pair survives at $\mu=0.7$, but it forms only a shallow sub-threshold structure rather than a conventional barrier above the continuum threshold. Indeed, solving simultaneously $\widetilde V'(x_M)=0$ and $\widetilde V(x_M)=\mu^2$ gives the lower crossover $\mu_b\simeq 0.63785$ at which the local maximum drops below the asymptotic plateau. The extrema themselves merge only at $\mu=\mu_c$.} The corresponding allowed region in the $(\mu,L)$--plane defined by Eq.~\eqref{muineq} is displayed in Fig.~\ref{fig:Ineq}, where the shaded area indicates the values of $\mu$ and $L$ for which the potential develops a local maximum followed by a local minimum. There, the blue diamond corresponds to the example $\mu=0.5<\mu_c(L=6)$.

\par \textcolor{black}{To relate the dimensionless parameter used in the spectral problem to a particle mass, we restore SI units}
\begin{equation}\label{eq:physical_mu}
\mu=\frac{G M_{\rm BH}}{c^{2}}\frac{m_{\rm phys} c}{\hbar}=\frac{r_{s}}{2\lambda_{C}} \;,
\end{equation}
where $r_{s}=2GM_{\rm BH}/c^{2}$ is the Schwarzschild radius and $\lambda_{C}=\hbar/(m_{\rm phys}c)$ is the reduced Compton wavelength. Thus, $\mu=GM_{\rm BH}m_{\rm phys}/(\hbar c)$ is the dimensionless gravitational coupling of the field to the black hole. Since $\lambda_C/r_s=(2\mu)^{-1}$, values $\mu\sim\mathcal{O}(1)$ correspond, for astrophysical black holes, to ultralight particles whose Compton wavelength is comparable to the horizon scale. \textcolor{black}{Conversely, $\mu\gg 1$ means only that the Compton wavelength is short relative to $r_s$. It does not by itself define the angular eikonal limit, which requires $\ell\gg 1$, nor does it guarantee an oscillatory far field, because the asymptotic wavenumber $k(\Omega)=\sqrt{\Omega^2-\mu^2}$ may be small or imaginary. At fixed low $\ell$, the term $\mu^2(1-1/x)$ dominates Eq.~\eqref{eq:VxpotM2}. The photon-sphere barrier can then disappear, as quantified by Eqs.~\eqref{root} and \eqref{muineq}. Numerically, Eq.~\eqref{eq:physical_mu} gives}
\begin{equation}\label{eq:mass_conversion}
\mu \simeq 7.48\times 10^{9}\left(\frac{M_{\rm BH}}{M_\odot}\right)\left(\frac{m_{\rm phys}}{\mathrm{eV}}\right) \;,\qquad
\textcolor{black}{m_{\rm phys}\simeq1.34\times10^{-10}\,\mu\left(\frac{M_\odot}{M_{\rm BH}}\right)\mathrm{eV}.}
\end{equation}

\begin{table*}[t]
\centering
\caption{\textit{\textcolor{black}{Representative physical scalar masses corresponding to the dimensionless couplings used in this work, obtained from Eq.~\eqref{eq:mass_conversion}, for a stellar-mass black hole of $10 \, M_\odot$ and supermassive black holes of $10^{6}\,M_\odot$ and $10^{8}\,M_\odot$. The first row gives the physical-mass range corresponding to the smallest and largest couplings considered in the principal range studied numerically, $0.1\leq\mu\leq2.2$; the second isolates the reference value $\mu=1$; and the remaining rows correspond to the large-$\mu$ numerical stress tests of Appendix~\ref{app:A}. Values are rounded to two significant figures.}}}
\label{table:physical_mass_conversion}
\setlength\tabcolsep{0.1cm}
\def\arraystretch{1.75}
\begin{tabular}{@{}|c|c|c|c|@{}}
\hline
\textcolor{black}{Dimensionless coupling} & \textcolor{black}{$m_{\rm phys}\,[\mathrm{eV}]$ for $10\,M_\odot$} & \textcolor{black}{$m_{\rm phys}\,[\mathrm{eV}]$ for $10^{6}\,M_\odot$} & \textcolor{black}{$m_{\rm phys}\,[\mathrm{eV}]$ for $10^{8}\,M_\odot$} \\ [0.5ex]
\hline
\textcolor{black}{$0.1\leq\mu\leq2.2$} & \textcolor{black}{$1.3\times10^{-12}\text{--}2.9\times10^{-11}$} & \textcolor{black}{$1.3\times10^{-17}\text{--}2.9\times10^{-16}$} & \textcolor{black}{$1.3\times10^{-19}\text{--}2.9\times10^{-18}$} \\
\hline
\textcolor{black}{$\mu=1$} & \textcolor{black}{$1.3\times10^{-11}$} & \textcolor{black}{$1.3\times10^{-16}$} & \textcolor{black}{$1.3\times10^{-18}$} \\
\hline
\textcolor{black}{$\mu=3$} & \textcolor{black}{$4.0\times10^{-11}$} & \textcolor{black}{$4.0\times10^{-16}$} & \textcolor{black}{$4.0\times10^{-18}$} \\
\hline
\textcolor{black}{$\mu=10$} & \textcolor{black}{$1.3\times10^{-10}$} & \textcolor{black}{$1.3\times10^{-15}$} & \textcolor{black}{$1.3\times10^{-17}$} \\
\hline
\textcolor{black}{$\mu=100$} & \textcolor{black}{$1.3\times10^{-9}$} & \textcolor{black}{$1.3\times10^{-14}$} & \textcolor{black}{$1.3\times10^{-16}$} \\ [0.5ex]
\hline
\end{tabular}
\end{table*}
\textcolor{black}{All of the masses in Table~\ref{table:physical_mass_conversion} are ultralight in natural units. Their phenomenological status is nevertheless model- and source-dependent. Although the present calculation concerns a test scalar field on a non-rotating Schwarzschild background, the most developed astrophysical constraints on ultralight bosons arise from rotating Kerr black holes, where the superradiant instability can amplify bosonic fields and deplete the black hole spin. It is therefore useful to compare the physical-mass ranges in Table~\ref{table:physical_mass_conversion} with representative superradiance bounds, while emphasising that such bounds do not apply directly to the Schwarzschild spectral problem studied here. For Kerr black holes, superradiance is most efficient when the analogous coupling $\alpha=GM_{\rm BH}m_{\rm phys}/(\hbar c)$ is of order a few tenths. Bounds inferred from spin measurements depend on the black hole spin and evolution time, accretion and companion effects, the superradiant levels included, and any boson self-interactions or non-gravitational couplings \cite{BritoCardosoPani2020_Superradiance,WitteMummery2025PRD,Hoof2026MNRAS}.}

\textcolor{black}{A recent GWTC-5 population analysis reports no evidence for superradiant axions and, within its axion and binary-population model, excludes $1.7\times10^{-14}\,\mathrm{eV}\lesssim m_a\lesssim3.3\times10^{-12}\,\mathrm{eV}$ at $95\%$ confidence \cite{NingSafdiWelch2026GWTC5}. For a $10\,M_\odot$ black hole, the physical masses corresponding to the main dimensionless range studied in our numerical analysis, $0.1\leq\mu\leq2.2$, overlap this particular excluded band only over approximately $1.3\times10^{-12}$--$3.3\times10^{-12}\,\mathrm{eV}$, corresponding to $0.10\lesssim\mu\lesssim0.25$. The fact that the cases $\mu>1$ lie above this particular band does not establish their generic viability. At supermassive black hole scales, a representative spin-based analysis reported sensitivity across approximately $10^{-22}$--$10^{-17}\,\mathrm{eV}$ and, for self-interacting scalar models with $f_a>10^{15}\,\mathrm{GeV}$, excluded most of $3\times10^{-19}$--$10^{-17}\,\mathrm{eV}$ \cite{ZuFengYuanFan2020EPJP}. In that specific model, the physical-mass range corresponding to $0.1\leq\mu\leq2.2$ for a $10^8\,M_\odot$ black hole overlaps the quoted interval over approximately $3\times10^{-19}$--$2.9\times10^{-18}\,\mathrm{eV}$, whereas the corresponding range for a $10^6\,M_\odot$ black hole lies mostly above it. More recent Bayesian work using the full mass--spin posterior of IRAS~09149--6206 shows explicitly that the quantitative inference changes with the black hole evolution time, the included levels, scalar self-interactions, and the treatment of observational uncertainties \cite{Hoof2026MNRAS}. A direct LIGO--Virgo--KAGRA O3 search likewise found no evidence for continuous gravitational waves from scalar clouds and set upper limits whose translation into a boson-mass exclusion depends, among other inputs, on the source and cloud age \cite{LVK2022ScalarCloud}. Hence, neither generic viability nor generic exclusion follows from $\mu$ alone. The present Schwarzschild test-field calculation assumes no cosmological abundance, production history, self-interaction, or detector coupling. Here, $\mu$ is primarily a control parameter for the spectral problem.}

\section{Extended spectral method \label{sec:method}}

Before we consider the massive Schwarzschild problem, it is useful to recall the logic of the spectral construction used in the earlier formulation of the method \cite{Batic2024CQG, Batic2024PRD, Batic2025EPJC, Batic2025CQG, Batic2025PRSA, Batic2026EPJC}. After factoring out the known QNM asymptotics at the two boundaries and compactifying the radial domain to the finite interval $y\in[-1,1]$, one is left with a regular unknown function, which is approximated by a truncated Chebyshev expansion,
\begin{equation}\label{eq:chebexp}
\Phi_N(y)=\sum_{k=0}^{N} c_k T_k(y).
\end{equation}
\textcolor{black}{Collocating the differential equation at the Gauss--Lobatto points produces a finite-dimensional polynomial eigenvalue problem for the coefficient vector $\mathbf{c}=(c_0,\dots,c_N)^{\top}$.} In the standard massless setting, the dependence on the frequency is a polynomial of degree two, so the discretised problem takes the form \eqref{eq:oldpencil}. \textcolor{black}{The earlier formulation therefore leads to a quadratic matrix pencil \cite{TisseurMeerbergen,Beyn2014}. Since the exact QNMs are not known \emph{a priori}, the numerical uncertainty of a spectral calculation is assessed \textit{a posteriori} through stability under refinement.} More precisely, if $\Omega_j^{(N)}$ denotes a mode computed with $N$ basis functions, and $\Omega_j^{(N+\Delta N)}$ is the corresponding mode obtained after increasing the truncation order, we use the difference between these two values as an estimate of the error
\begin{equation}\label{eq:spectral-error}
\varepsilon_j^{(N)}=\left|\Omega_j^{(N+\Delta N)}-\Omega_j^{(N)}\right|.
\end{equation}
\textcolor{black}{For the calculations reported below, refinement is performed simultaneously in truncation order and arithmetic precision. For each parameter pair we compute the complete spectra at $(N,p)\in\{(180,180),(190,190),(200,200)\}$, where $p$ denotes the decimal working precision. A root is retained only if it can be matched across all three spectra and satisfies}
\begin{equation}\label{eq:threeway-error}
\textcolor{black}{\delta_j=\max_{N,N'\in\{180,190,200\}}\left|\Omega_j^{(N)}-\Omega_j^{(N')}\right|\leq 10^{-4}.}
\end{equation}
\textcolor{black}{Throughout the paper, the term \emph{stable spectral root} refers precisely to a root passing the three-way resolution test and the increased-precision check. This convergence criterion does not, by itself, assign a Riemann sheet or a far-field physical class. Strongly drifting roots are discarded. The raw $\Omega$ spectra and the convergence reports used to construct the tables are publicly available with the numerical code in Ref. \cite{Batic2026_MassiveScalar}.}

\par \textcolor{black}{The massive case preserves this strategy but changes the algebraic structure of the eigenvalue problem. As introduced in Section~\ref{sec:intro}, the asymptotic wavenumber is the multivalued quantity $k(\Omega)=\sqrt{\Omega^2-\mu^2}$, so the discretised operator is not a polynomial in $\Omega$. The ESM restores polynomiality by introducing a uniformising spectral parameter. The quadratic pencil of the massless formulation is thereby replaced by a seventh-degree polynomial pencil. This is not a different numerical philosophy, but a direct generalisation of the earlier construction to a problem with nontrivial branch structure in the frequency plane. The resulting degree-seven matrix polynomial is solved with the companion linearisation implemented with MATLAB's \texttt{polyeig} routine. For a regular matrix polynomial, this is a strong linearisation because its finite eigenvalues and their algebraic multiplicities are those of the discretised polynomial problem itself \cite{TisseurMeerbergen}. Linearisation can affect conditioning, but it does not define an additional finite spectrum. Consequently, an eigenvalue that reappears after the differential operator has been reassembled at independent values of $N$ cannot be attributed solely to the seventh-degree companion form. This point is corroborated by the recent analysis of Ref.~\cite{BaticDutykhSukaiti2026KSQC}, where analogous purely imaginary ladders were obtained in four perturbation sectors from a quadratic matrix pencil. That work also performed a wide-resolution drift test over central resolutions $125\leq N\leq260$: fixed imaginary-axis levels formed nearly horizontal plateaus, while increasing $N$ extended the branch to larger damping instead of moving previously resolved roots. Although that calculation concerns a different black hole family and is not a frequency benchmark for the present Schwarzschild modes, it {\color{black}{provides suggestive structural evidence}} that this type of imaginary-axis structure is neither tied to a degree-seven companion pencil nor confined to a narrow resolution triplet. With this overview in place, we return to Eq.~\eqref{radial}, which can be written as}
\begin{equation}\label{ODE1}
  \frac{d^2\psi}{dx^2}+\mathfrak{p}(x)\frac{d\psi}{dx}+\mathfrak{q}(x)\psi(x)=0   
\end{equation}
with 
\begin{equation}\label{pq}
\mathfrak{p}(x)=-\frac{1}{x}+\frac{1}{x-1},\quad 
\mathfrak{q}(x)=\left(\frac{2x}{x-1}\right)^2\left[\Omega^2-\mu^2+\frac{\mu^2}{x}-\frac{L}{4x^2}+\frac{L-1}{4x^3}+\frac{1}{4x^4}\right].
\end{equation}
Because $\mathfrak{p}(x)$ and $\mathfrak{q}(x)$ have poles of order one and two at $x=1$, respectively, this point is a regular singular point of Eq.~\eqref{ODE1}. Frobenius theory \cite{Ince1956ordinary} therefore leads us to seek solutions of the form
\begin{equation}
\psi(x) = (x-1)^\rho\sum_{\kappa=0}^\infty a_\kappa(x-1)^\kappa.
\end{equation}
The leading behaviour at $x=1$ is represented by $(x-1)^\rho$, where $\rho$ is determined by the indicial equation
\begin{equation}\label{indicial}
\rho(\rho-1) + \mathfrak{p}_0\rho + \mathfrak{q}_0 = 0,
\end{equation}
with
\begin{equation}
\mathfrak{p}_0=\lim_{x \to 1}(x-1)\mathfrak{p}(x) = 1,\qquad
\mathfrak{q}_0=\lim_{x \to 1}(x-1)^2 \mathfrak{q}(x)=4\Omega^2.
\end{equation}
The roots of Eq.~\eqref{indicial} are $\rho_\pm=\pm2i\Omega$, and the QNM boundary condition at the event horizon $x=1$ is
\begin{equation}\label{QNMBCz1}
\psi\underset{{x\to 1^+}}{\longrightarrow} (x-1)^{-2i\Omega}.
\end{equation}
\textcolor{black}{With the convention $e^{-i\omega t}$, this factor is proportional to $e^{-i\omega r_*}$ near the horizon (where the Schwarzschild tortoise coordinate is defined by $dr_*/dr=1/f(r)$) and hence gives the ingoing dependence $e^{-i\omega(t+r_*)}$ on the advanced null coordinate $v=t+r_*$. The asymptotic behaviour of the solutions to Eq.~\eqref{ODE1} can be obtained by the method of Ref.~\cite{Olver1994MAA}. As $x\to+\infty$,}
\begin{equation}\label{pqS}
\mathfrak{p}(x)=\sum_{\kappa=0}^\infty\frac{\mathfrak{f}_\kappa}{x^\kappa}=\mathcal{O}\left(\frac{1}{x^2}\right),\qquad
\mathfrak{q}(x)=\sum_{\kappa=0}^\infty\frac{\mathfrak{g}_\kappa}{x^\kappa}=4(\Omega^2-\mu^2)+\frac{8\Omega^2-4\mu^2}{x}+\mathcal{O}\left(\frac{1}{x^2}\right) \;.
\end{equation}
\textcolor{black}{Given that at least one of the coefficients $\mathfrak{f}_0$, $\mathfrak{g}_0$, and $\mathfrak{g}_1$ is nonzero, a formal asymptotic solution to Eq.~\eqref{ODE1} is represented by \cite{Olver1994MAA}}
\begin{equation}\label{olvers}
\textcolor{black}{\psi^{(\pm)}(x)=x^{\nu_\pm}e^{\sigma_\pm x}\sum_{\kappa=0}^\infty\frac{a_{\kappa,\pm}}{x^\kappa} \;.}
\end{equation}
\textcolor{black}{Here, $\sigma_\pm$ are the exponential characteristic exponents and $\nu_\pm$ are the corresponding power-law exponents, determined by}
\begin{equation}\label{chareqns}
\textcolor{black}{\sigma_\pm^2+\mathfrak{f}_0\sigma_\pm+\mathfrak{g}_0=0 \;,\qquad
\nu_\pm=-\frac{\mathfrak{f}_1\sigma_\pm+\mathfrak{g}_1}{\mathfrak{f}_0+2\sigma_\pm} \;.}
\end{equation}
\textcolor{black}{Using the asymptotic wavenumber $k$ and the Coulomb parameter $\eta$,}
\begin{equation}\label{eq:keta}
\textcolor{black}{k=\sqrt{\Omega^2-\mu^2} \;,\qquad
\eta=\frac{2\Omega^2-\mu^2}{k} \;,}
\end{equation}
\textcolor{black}{respectively, one obtains $\sigma_\pm=\pm 2ik$, and $\nu_\pm=\pm i\eta$. For the time dependence $e^{-i\omega t}$, the outgoing solution is the plus branch, analytically continued from $k\sim+\Omega$ at large $|\Omega|$. Its spatial-infinity boundary condition is therefore}
\begin{equation}\label{QNMBCposinf}
\textcolor{black}{\psi_{\rm out}(x)\underset{x\to+\infty}{\sim}x^{i\eta}e^{2ikx}=x^{i\eta}e^{i\gamma x},\qquad \gamma=2k \;.}
\end{equation}
\textcolor{black}{Notice that the factors $i$ are already contained in $\sigma_+=2ik$ and $\nu_+=i\eta$. To apply the spectral method, we introduce a new radial function $\Phi(x)$ such that the boundary conditions in Eqs.~\eqref{QNMBCz1} and \eqref{QNMBCposinf} are built into the ansatz and $\Phi(x)$ is regular at the event horizon and at spatial infinity}
\begin{eqnarray}
\psi(x)&=&x^{i\alpha}(x-1)^{-i\beta}e^{i\gamma(x-1)}\Phi(x)\; ,\label{AnsatzS}\\
\alpha&=&2\Omega+\frac{2\Omega^2-\mu^2}{\sqrt{\Omega^2-\mu^2}} \;,\quad
\beta=2\Omega \;,\quad
\gamma=2\sqrt{\Omega^2-\mu^2} \;,\quad\Omega\neq \pm\mu \;.
\end{eqnarray}
\textcolor{black}{Indeed, $\alpha-\beta=\eta$ and $\gamma=2k$, so Eq.~\eqref{AnsatzS} has the large-$x$ behaviour $x^{i\eta}e^{2ikx}\Phi(x)$ and is exactly consistent with Eq.~\eqref{QNMBCposinf}. Substituting Eq.~\eqref{AnsatzS} into Eq.~\eqref{ODE1} yields the following ordinary differential equation for $\Phi$}
\begin{equation}\label{ODEznone}
P_2(x)\Phi^{''}(x)+P_1(x)\Phi^{'}(x)+P_0(x)\Phi(x)=0 \;,
\end{equation}
with
\begin{eqnarray}
P_2(x)&=&1 \;,\\
P_1(x)&=&2i\gamma+\frac{2i\alpha-1}{x}+\frac{1-2i\beta}{x-1} \;,\\
P_0(x)&=&\frac{L+1-2\alpha(\beta+\gamma)-i(\alpha+\beta+\gamma)}{x}
+\frac{(\beta+\gamma)^2+2\alpha\beta-L-1+i(\alpha+\beta+\gamma)}{x-1}
+\frac{(1-i\alpha)^2}{x^2} \;.
\end{eqnarray}
\textcolor{black}{We next introduce $x=2/(1-y)$, which maps spatial infinity and the event horizon to $y=1$ and $y=-1$, respectively. Eq.~\eqref{ODEznone} then becomes}
\begin{equation}\label{ODEynone}
S_2(y)\ddot{\Phi}(y)+S_1(y)\dot{\Phi}(y)+S_0(y)\Phi(y)=0 \;,
\end{equation}
where a dot denotes differentiation with respect to the new variable $y$, and
\begin{align}
S_2(y)&=\frac{(1-y)^4}{4} \;,\label{S2}\\
S_1(y)&=-\frac{(1-y)^3}{2}+\frac{(1-y)^2}{2}\left[2i\gamma+(1-2i\beta)\frac{1-y}{1+y}+\frac{2i\alpha-1}{2}(1-y)\right],\label{S1}\\
S_0(y)&=\frac{1}{2}\left[L+1-2\alpha(\beta+\gamma)-i(\alpha+\beta+\gamma)\right](1-y)+
\left[(\beta+\gamma)^2+2\alpha\beta-L-1+i(\alpha+\beta+\gamma)\right]\frac{1-y}{1+y}\nonumber\\
&+\frac{(1-i\alpha)^2}{4}(1-y)^2 \;.\label{S0}
\end{align}
\textcolor{black}{Regularity of $\Phi(y)$ at $y=\pm1$ is also required.}
\begin{table}[t]
\caption{\textit{Classification of the points $y=\pm1$ for the functions defined in Eqs.~\eqref{S2}--\eqref{S0}.}}
\begin{center}
\setlength\tabcolsep{0.1cm}
\def\arraystretch{1.5}
\begin{tabular}{@{} | c | c | c | c | c | c | c | c @{}}
\hline
$y$  & $S_2(y)$       & $S_1(y)$       & $S_0(y)$\\ \hline
$-1$ & +4              & \mbox{pole of order} 1 & \mbox{pole of order} 1\\ \hline
$+1$ & \mbox{zero of order} 4 & \mbox{zero of order} 2 & \mbox{zero of order} 2\\ \hline
\end{tabular}
\label{tableEinsnone}
\end{center}
\end{table}
Table~\ref{tableEinsnone} shows that the coefficients in Eq.~\eqref{ODEynone} share a common zero of order $2$ at $y=1$, whereas $S_1$ and $S_0$ have a pole of order $1$ at $y=-1$. To obtain a form suitable for the spectral method, we multiply Eq.~\eqref{ODEynone} by $(1+y)/(1-y)^2$ and use the identities
\begin{equation}\label{ident}
\beta^2-\gamma^2-4\mu^2=0,\quad
(\alpha-\beta)\gamma-\frac{\beta^2+\gamma^2}{2}=0 \;.
\end{equation}
Eq.~\eqref{ODEynone} then reduces to
\begin{equation}\label{ODEhynone}
M_2(y)\ddot{\Phi}(y)+M_1(y)\dot{\Phi}(y)+M_0(y)\Phi(y)=0,
\end{equation}
where
\begin{align}
M_2(y)&=\frac{(1+y)(1-y)^2}{4} \;, \\
M_1(y)&=i\gamma(1+y)+\frac{1-2i\beta}{2}(1-y)+\frac{1}{4}\left[i\frac{(\beta+\gamma)^2}{\gamma}-3\right](1-y^2) \;, \\
M_0(y)&=-\frac{L+1}{2}+\frac{(\beta+\gamma)^3}{2\gamma}
+i\frac{(\beta+\gamma)(\beta+3\gamma)}{4\gamma}
+\frac{1}{4}\left[1-\frac{(\beta+\gamma)^2}{2\gamma}\right]^2(1+y) \;.
\end{align}
The corresponding endpoint limits are
\begin{align}
&\lim_{y\to 1^{-}}M_2(y)=0=\lim_{y\to -1^{+}}M_2(y) \;,\\
&\lim_{y\to 1^{-}}M_1(y)=2i\gamma,\quad
\lim_{y\to -1^{+}}M_1(y)=1-2i\beta \;,\\
&\lim_{y\to 1^{-}}M_0(y)=-\frac{L}{2}-\frac{(\beta+\gamma)^3(\beta-3\gamma)}{8\gamma^2}+i\frac{\gamma^2-\beta^2}{4\gamma} \;,\\
&\lim_{y\to -1^{+}}M_0(y)=-\frac{L+1}{2}+\frac{(\beta+\gamma)^3}{2\gamma}+i\frac{(\beta+\gamma)(\beta+3\gamma)}{4\gamma} \;.
\end{align}
\begin{table}[t]
\caption{\textit{Definitions of the coefficients $L_{kj}$ appearing in Eq.~\eqref{Lkj}, which depend on the mass parameter $\mu$ and the angular-momentum parametrisation $L=\ell(\ell+1)$, together with their endpoint behaviour on $-1\leq y\leq1$.}}
\begin{center}
\setlength\tabcolsep{0.1cm}
\def\arraystretch{1.5}
\begin{tabular}{@{} | c | c | c | c | c | c | c | c @{}}
\hline
$(k,j)$  & $\displaystyle{\lim_{y\to -1^+} L_{kj}}$  & $L_{kj}$ & $\displaystyle{\lim_{y\to 1^-}L_{kj}}$  \\[2ex] \hline
$(0,0)$ &  $0$          & $0$             & $0$\\ 
$(0,1)$ &  $-8\mu^6$    & $-8\mu^6$       & $-8\mu^6$\\ 
$(0,2)$ &  $0$          & $0$             & $0$\\ \hline 
$(1,0)$ &  $2\mu^4(L+1)$& $\mu^4(2L+1-y)$ & $2\mu^4 L$\\ 
$(1,1)$ &  $-4\mu^4$    & $-\mu^4(3y^2-2y-1)$& $0$\\ 
$(1,2)$ &  $0$          & $-\mu^4(1+y)(1-y)^2$  & $0$\\ \hline 
$(2,0)$ &  $4\mu^4$     & $4\mu^4$        & $4\mu^4$\\ 
$(2,1)$ &  $8\mu^4$     & $8\mu^4(y+2)$   & $24\mu^4$\\ 
$(2,2)$ &  $0$          & $0$             & $0$\\ \hline
$(3,0)$ &  $-4\mu^2(L+1)$  & $-2\mu^2(2L+1-y)$ & $-4\mu^2 L$\\ 
$(3,1)$ &  $8\mu^2$     & $2\mu^2(3y^2-2y-1)$ & $0$\\ 
$(3,2)$ &  $0$          & $2\mu^2(1+y)(1-y)^2$& $0$\\ \hline
$(4,0)$ &  $-12\mu^2$   & $4\mu^2(y-2)$   & $-4\mu^2$\\ 
$(4,1)$ &  $8\mu^2$     & $4\mu^2(y^2-4y-3)$ & $-24\mu^2$\\ 
$(4,2)$ &  $0$          & $0$& $0$\\ \hline
$(5,0)$ &  $16\mu^2+2(L+1)$   & $16\mu^2+2L+1-y$   & $8(2\mu^2+L)$\\ 
$(5,1)$ &  $-4$     & $-(3y^2-2y-1)$ & $0$\\ 
$(5,2)$ &  $0$          & $-(1+y)(1-y)^2$& $0$\\ \hline
$(6,0)$ &  $8$    & $4(1-y)$   & $0$\\ 
$(6,1)$ &  $-8$   & $-4(y^2-2y-1)$ & $8$\\ 
$(6,2)$ &  $0$          & $0$& $0$\\ \hline
$(7,0)$ &  $-16$ & $4(y-3)$   & $-8$\\ 
$(7,1)$ &  $0$   & $0$ & $0$\\ 
$(7,2)$ &  $0$   & $0$ & $0$\\ \hline
\end{tabular}
\label{tableZweinone}
\end{center}
\end{table} 
Because $\alpha$ and $\gamma$ depend explicitly on the non-polynomial quantity $\sqrt{\Omega^2-\mu^2}$, discretising Eq.~\eqref{ODEynone} does not directly yield a polynomial eigenvalue problem in $\Omega$. To recover polynomial dependence, we introduce the transformation,
\begin{equation}\label{trafo}
\Lambda=\Omega+\sqrt{\Omega^{2}-\mu^{2}}.
\end{equation}
This is a scaled form of the classical Joukowski conformal map, $z\mapsto z+a^2/z$, introduced in airfoil theory by \textcolor{black}{Joukowski} \cite{Joukowsky1910ZFM} and discussed in modern treatments of conformal mapping \cite{Schinzinger1991}. \textcolor{black}{As explained in Section~\ref{sec:intro}, the map uniformises the multivalued dispersion relation. Eq.~\eqref{trafo} gives}
\begin{equation}\label{Omegaform}
\Omega=\frac{\Lambda^2+\mu^2}{2\Lambda} \;,\quad
\sqrt{\Omega^{2}-\mu^{2}}=\frac{\Lambda^2-\mu^2}{2\Lambda} \;.
\end{equation}
\textcolor{black}{A straightforward calculation gives the corresponding parameters in Eq.~\eqref{AnsatzS}}
\begin{equation}
\alpha=\frac{2\Lambda^3}{\Lambda^2-\mu^2} \;,\quad
\beta=\frac{\Lambda^2+\mu^2}{\Lambda} \;,\quad
\gamma=\frac{\Lambda^2-\mu^2}{\Lambda} \;.
\end{equation}
\textcolor{black}{The ansatz in Eq.~\eqref{AnsatzS} incorporates the QNM boundary conditions and reproduces the required far-field behaviour. In particular, since}
\begin{equation}
\alpha-\beta=\frac{2\Omega^2-\mu^2}{\sqrt{\Omega^2-\mu^2}} \;,\qquad
\gamma=2\sqrt{\Omega^2-\mu^2} \;,
\end{equation}
\textcolor{black}{the large-$x$ behaviour is $\psi(x)\sim x^{\,i(\alpha-\beta)}e^{i\gamma x}$, up to an $x$-independent factor and inverse-power corrections. The asymptotic solution therefore has the expected Coulomb-type power-law prefactor multiplied by $e^{i\gamma x}$.} The sign of $ \sqrt{\Omega^2-\mu^2}$, equivalently of the asymptotic wavenumber, $k(\Omega)=\sqrt{\Omega^2-\mu^2}=\gamma/2$, is therefore part of the boundary condition at spatial infinity. \textcolor{black}{For a real frequency, a real $k$ gives an oscillatory far field and a purely imaginary $k$ gives an exponentially varying one. For a complex frequency, $k$ is generically complex, so the labels \emph{propagating} and \emph{evanescent} are not mutually exclusive physical classes. The outgoing sheet is defined by analytic continuation from $k(\Omega)\sim+\Omega$ as $|\Omega|\to\infty$. The reciprocal sheet has the opposite sign of $k$.} The Joukowski transformation of Eq.~\eqref{trafo} uniformises the corresponding two-sheeted Riemann surface. Using Eq.~\eqref{Omegaform}, we may write
\begin{equation}
k(\Lambda)=\frac{\Lambda^2-\mu^2}{2\Lambda} \;.
\end{equation}
Hence, the two points on the Riemann surface lying over the same value of $\Omega$ are represented by the reciprocal pair
\begin{equation}
\Lambda \longleftrightarrow \frac{\mu^2}{\Lambda} \;,
\end{equation}
under which $\Omega$ is unchanged, whereas $k$ and $\gamma$ change sign. Equivalently, the branch points $\Omega=\pm\mu$ are mapped to $\Lambda=\pm\mu$, and the two sheets are separated in the $\Lambda$-plane by the circle $|\Lambda|=\mu$. The branch determined by $k(\Omega)\sim +\Omega$ corresponds to the exterior region $|\Lambda|>\mu$, while the reciprocal interior region $|\Lambda|<\mu$ represents the opposite sheet. Accordingly, the polynomial eigenvalue problem in $\Lambda$ may return algebraic roots associated with either sheet. \textcolor{black}{If we let $k=k_{\rm R}+ik_{\rm I}$ so that $e^{i\gamma x}=e^{2ikx}=e^{2ik_{\rm R}x}e^{-2k_{\rm I}x}$, then $k_{\rm R}\neq 0$ produces an oscillatory radial phase, $k_{\rm I}>0$ produces exponential decay at infinity, and $k_{\rm I}<0$ gives rise to exponential spatial growth, up to the Coulomb-type power-law prefactor. In particular, for a damped purely imaginary frequency $\Omega=-i\Gamma$ with $\Gamma>0$, the two sheets give $k_{\rm out}=-i\sqrt{\Gamma^2+\mu^2}$ and $k_{\rm rec}=+i\sqrt{\Gamma^2+\mu^2}$. The first is the spatially growing analytic continuation of the outgoing resonance condition, whereas the reciprocal value is exponentially decaying. Both have the same $\Omega$ and the same $\delta_{\rm R}=-\mu^2$. Hence, a purely imaginary frequency cannot be classified from $\Omega$ or $\delta_{\rm R}$ alone. A sheet-resolved post-processing must retain $\Lambda$, because the reciprocal pair $\Lambda$ and $\mu^2/\Lambda$ maps to the same $\Omega$ while reversing $k$. The archived files used to assemble the present tables retain only the mapped frequency $\Omega$. Consequently, the sheet cannot be reconstructed unambiguously from those files alone. The output routine writes $\Lambda$, $\Omega$, $k(\Lambda)$, the sheet indicator $\operatorname{sgn}(|\Lambda|-\mu)$, and the exponential indicator $\operatorname{sgn}(\operatorname{\Im {\rm m}}k)$ in addition to the original two-column $\Omega$ file. A fully sheet-resolved re-tabulation requires rerunning this output step. Until then, the displayed groups are distinguished only by the neutral proxy $\delta_{\rm R}$. Finally, the Joukowski transformation in Eq.~\eqref{trafo} yields the polynomial spectral problem}
\begin{equation}\label{TSCH}
L_0+iL_1\Lambda+L_2\Lambda^2+iL_3\Lambda^3+L_4\Lambda^4+iL_5\Lambda^5+L_6\Lambda^6+iL_7\Lambda^7= 0
\end{equation}
with
\begin{equation}\label{Lkj}
L_k=L_{k0}(y)\Phi(y)+L_{k1}(y)\dot{\Phi}(y)+L_{k2}(y)\ddot{\Phi}(y)
\end{equation}
\textcolor{black}{for each $k\in\{0,1,\ldots,7\}$. Table~\ref{tableZweinone} summarises the coefficients $L_{kj}$ in Eq.~\eqref{Lkj} and their endpoint values at $y=\pm1$. Once $\Lambda$ has been determined numerically, $\Omega$ is recovered from the first relation in Eq.~\eqref{Omegaform}. Because $\Omega$ is a rational function of $\Lambda$, this recovery requires neither a square root nor an additional branch choice. The formulation above yields a well-defined polynomial eigenvalue problem of degree seven in the spectral parameter $\Lambda$, from which the frequency $\Omega$ is recovered through Eq.~\eqref{Omegaform}. This reformulation makes explicit that the increased algebraic degree is a direct consequence of uniformising the multivalued dispersion relation, rather than an artefact of the discretisation. In the following section, we compute stable massive scalar spectral roots and organise the existing $\Omega$-only data by the nominal proxy $\delta_{\rm R}$, without using that proxy as a sheet or asymptotic classification. The square-root branch associated with $k(\Omega)$ is not represented numerically by a row of points sampling an arbitrarily chosen cut. The two-sheeted surface is uniformised before Chebyshev discretisation. The degree-seven problem can nevertheless contain roots on both sheets. In the remainder of the paper, a \emph{purely imaginary {\color{black}{root}}} means a stable spectral root under the convergence criterion above. The present $\Omega$-only archive does not establish whether an individual tabulated root belongs to the outgoing or reciprocal sheet. This spectral statement is also distinct from the stronger question of the residue with which a pole enters a Green-function decomposition or a waveform generated by specified initial data. {\color{black}{Accordingly, we do not identify such imaginary roots as physical QNMs.}}}

\section{Numerical results \label{sec:results}}

\textcolor{black}{We first validate the ESM against the sixth-order WKB results of Ref.~\cite{Konoplya:2003ii} and the continued-fraction results of Ref.~\cite{Leaver1985PRSLA}. Setting $\mu=0$ in Eq.~\eqref{TSCH} reduces the master equation to the massless Schwarzschild problem. Table~\ref{tablemu0} shows the corresponding QNFs and reproduces the expected dependence on $\ell$ and $n$. At the reported precision, the spectral results obtained with $N=40$ and $N=200$ agree to order $10^{-4}$. This agreement validates the $\mu=0$ reduction and the common spectral assembly, but it does not by itself test the massive asymptotic factors, the branch prescription, or the uniformisation; we check these massive-specific ingredients in the next subsection.}
\begin{table}[t]
\centering
\caption{\textit{QNFs of massless scalar perturbations $(\mu=0)$ of the Schwarzschild black hole. The $N=200$ extended spectral results (final column) are compared with the continued-fraction values of Ref.~\cite{Leaver1985PRSLA} (third column), the sixth-order WKB values of Ref.~\cite{Konoplya:2003ii} (fourth column), and the $N=40$ spectral results of Ref.~\cite{Mamani2022EPJC}. Entries labelled \enquote{N/A} indicate unavailable data.}}
\label{tablemu0}
\setlength\tabcolsep{0.1cm}
\def\arraystretch{1.5}
 \begin{tabular}{@{}|c| c| c| c| c| c|@{}} 
 \hline
 $\ell$ & $n$ & $\Omega_{\rm CFM}$ \cite{Leaver1985PRSLA} & $\Omega_{\rm WKB}$ \cite{Konoplya:2003ii} & $\Omega_{\rm N=40}$ \cite{Mamani2022EPJC} & $\Omega_{\rm N=200}$\\ [0.5ex] 
 \hline
 $0$ & $0$ & $0.1105-0.1049i$ & $0.1105-0.1008i$ & $0.1105-0.1049i$ & $0.1105-0.1049i$\\ 
     & $1$ & $0.0861-0.3481i$ & $0.0890-0.3446i$ & \mbox{N/A}       & $0.0861-0.3481i$\\
     & $2$ & \mbox{N/A}       & $0.1918-0.4765i$       & \mbox{N/A}       & $0.0757-0.6011i$\\
     \hline
 $1$ & $0$ & $0.2929-0.0977i$ & $0.2929-0.0978i$ & $0.2929-0.0977i$ & $0.2929-0.0977i$\\
     & $1$ & $0.2645-0.3063i$ & $0.2645-0.3065i$ & $0.2645-0.3063i$ & $0.2644-0.3063i$\\
     & $2$ & $0.2295-0.5401i$ & $0.2310-0.5422i$ & \mbox{N/A}       & $0.2295-0.5401i$\\ 
     & $3$ & $0.2033-0.7883i$ & $0.2221-0.7952i$ & \mbox{N/A}       & $0.2033-0.7883i$\\
     \hline
 $2$ & $0$ & $0.4836-0.0968i$ & $0.4836-0.0968i$ & $0.4836-0.0968i$ & $0.4836-0.0968i$\\
     & $1$ & $0.4639-0.2956i$ & $0.4638-0.2956i$ & $0.4639-0.2956i$ & $0.4639-0.2956i$\\
     & $2$ & $0.4305-0.5086i$ & $0.4304-0.5087i$ & $0.4305-0.5086i$ & $0.4305-0.5086i$\\
     & $3$ & $0.3939-0.7381i$ & $0.3932-0.7399i$ & \mbox{N/A}       & $0.3939-0.7381i$\\ [0.5ex] 
 \hline
 \end{tabular}
\end{table}

\subsection{\textcolor{black}{Massive-mode benchmarks}\label{sec:massive-benchmarks}}

\textcolor{black}{Because the novelty of the present calculation lies in the massive problem, the massless check in Table~\ref{tablemu0} is not used as the sole validation. We have also added a two-step massive-field comparison. First, we calibrated an independent implementation of the Leaver-Nollert CFM against the published Hill-determinant data of Ref.~\cite{AlvesPonquioMedeiros2025PRD}, whose study also cross-checks Hill determinants against a continued fraction. Matching the mass term in their horizon-scaled radial equation gives $m_{\rm Ref.}=2\mu$. The frequency comparison is made in the $M\omega$ normalisation fixed by the common massless benchmark. Their printed $m_{\rm Ref.}=0.1$ values therefore correspond to $\mu=0.05$ here. Table~\ref{table:literature-massive-benchmark} shows agreement at the level permitted by the digits printed in that reference. We then use the same independently calibrated recurrence at the exact values of $\mu$, $\ell$, and mode index $n$ appearing in our spectral tables, thereby avoiding the uncontrolled precision that would result from digitising published frequency trajectories.}

\begin{table}[H]
\centering
\caption{\textit{\textcolor{black}{Calibration of the independent continued-fraction implementation against the published massive scalar Hill-determinant frequencies of Ref.~\cite{AlvesPonquioMedeiros2025PRD}. After matching the mass term, the reference value $m_{\rm Ref.}=0.1$ corresponds to $\mu=0.05$ in the present convention. The final column is the absolute complex-frequency difference. It is limited by the number of digits printed in the reference.}}}
\label{table:literature-massive-benchmark}
\setlength\tabcolsep{0.1cm}
\def\arraystretch{1.5}
\begin{tabular}{@{}|c|c|c|c|c|@{}}
\hline
\textcolor{black}{$\ell$} & \textcolor{black}{$n$} & \textcolor{black}{$\Omega_{\rm CFM}$ (this check)} & \textcolor{black}{$\Omega_{\rm Hill}$ \cite{AlvesPonquioMedeiros2025PRD}} & \textcolor{black}{$|\Delta\Omega|$} \\ [0.5ex]
\hline
\textcolor{black}{$0$} & \textcolor{black}{$0$} & \textcolor{black}{$0.1109947833-0.1028458207i$} & \textcolor{black}{$0.110988-0.102842i$} & \textcolor{black}{$7.8\times10^{-6}$}\\
\textcolor{black}{$1$} & \textcolor{black}{$0$} & \textcolor{black}{$0.2940543157-0.0969879789i$} & \textcolor{black}{$0.294054-0.096988i$} & \textcolor{black}{$3.2\times10^{-7}$}\\
\textcolor{black}{$2$} & \textcolor{black}{$0$} & \textcolor{black}{$0.4844331754-0.0964882240i$} & \textcolor{black}{$0.484433-0.0964882i$} & \textcolor{black}{$1.8\times10^{-7}$} \\ [0.5ex]
\hline
\end{tabular}
\end{table}

\textcolor{black}{Second, we evaluated the three-term massive scalar recurrence and the Nollert remainder of Konoplya and Zhidenko \cite{Konoplya2005PLB} independently of the Chebyshev matrices. The archived spectral values were used only as starting guesses for locating the nearby roots. The continued-fraction residual, branch choice, and converged frequencies are determined entirely by the published recurrence. The calculation used $80$-digit arithmetic and truncations $N_{\rm CFM}\in\{300,500,800,1200\}$. The slowly converging $\mu=0.2$, $\ell=0$ fundamental mode was followed further through $N_{\rm CFM}\in\{1200,2000,3000,5000\}$, with a final $3000\to 5000$ change of $6.4\times10^{-16}$. For the outgoing branch, we impose the analytic continuation $k\sim+\Omega$ at large $|\Omega|$. For a spatially decaying check, we choose $\Im {\rm m}\{k\}>0$ in the asymptotic factor $e^{2ikx}$. Table~\ref{table:direct-massive-benchmark} compares this recurrence calculation directly with the archived spectral frequencies underlying the manuscript tables. The symbol $j$ orders the selected low-damping roots within the decaying sequence and is not asserted to be a global QNM overtone number.}

\begin{table*}[t]
\centering
\caption{\textit{\textcolor{black}{Direct massive-field comparison between the ESM and an independent Leaver-Nollert CFM based on Ref.~\cite{Konoplya2005PLB}. The label ``out'' denotes analytic continuation from $k\sim+\Omega$, while ``dec'' denotes the spatially decaying choice $\Im {\rm m}\{k\}>0$ in $e^{2ikx}$. The ESM column contains the archived high-precision frequencies underlying the tables.}}}
\label{table:direct-massive-benchmark}
\setlength\tabcolsep{0.1cm}
\def\arraystretch{1.5}
\begin{tabular}{@{}|c|c|c|c|c|c|c|@{}}
\hline
\textcolor{black}{$\mu$} & \textcolor{black}{$\ell$} & \textcolor{black}{index} & \textcolor{black}{branch} & \textcolor{black}{$\Omega_{\rm ESM}$} & \textcolor{black}{$\Omega_{\rm CFM}$} & \textcolor{black}{$|\Delta\Omega|$} \\ [0.5ex]
\hline
\textcolor{black}{$0.1$} & \textcolor{black}{$0$} & \textcolor{black}{$n=0$} & \textcolor{black}{out} & \textcolor{black}{$0.112361056497-0.096823028936i$} & \textcolor{black}{$0.112361056497-0.096823028936i$} & \textcolor{black}{$2.7\times10^{-15}$}\\
\textcolor{black}{$0.1$} & \textcolor{black}{$1$} & \textcolor{black}{$n=0$} & \textcolor{black}{out} & \textcolor{black}{$0.297415661245-0.094957073606i$} & \textcolor{black}{$0.297415661245-0.094957073606i$} & \textcolor{black}{$1.8\times10^{-17}$}\\
\textcolor{black}{$0.1$} & \textcolor{black}{$1$} & \textcolor{black}{$n=1$} & \textcolor{black}{out} & \textcolor{black}{$0.264688550967-0.302850712420i$} & \textcolor{black}{$0.264688550967-0.302850712420i$} & \textcolor{black}{$1.4\times10^{-17}$}\\
\textcolor{black}{$0.1$} & \textcolor{black}{$2$} & \textcolor{black}{$n=0$} & \textcolor{black}{out} & \textcolor{black}{$0.486803751960-0.095674581470i$} & \textcolor{black}{$0.486803751960-0.095674581470i$} & \textcolor{black}{$6.0\times10^{-18}$}\\
\textcolor{black}{$0.1$} & \textcolor{black}{$2$} & \textcolor{black}{$n=2$} & \textcolor{black}{out} & \textcolor{black}{$0.430664774112-0.506464067063i$} & \textcolor{black}{$0.430664774112-0.506464067063i$} & \textcolor{black}{$5.2\times10^{-17}$}\\
\hline
\textcolor{black}{$0.2$} & \textcolor{black}{$0$} & \textcolor{black}{$n=0$} & \textcolor{black}{out} & \textcolor{black}{$0.116206680134-0.075358181322i$} & \textcolor{black}{$0.116206571044-0.075358287445i$} & \textcolor{black}{$1.5\times10^{-7}$}\\
\textcolor{black}{$0.2$} & \textcolor{black}{$1$} & \textcolor{black}{$n=0$} & \textcolor{black}{out} & \textcolor{black}{$0.310956908398-0.086593285616i$} & \textcolor{black}{$0.310956908398-0.086593285616i$} & \textcolor{black}{$2.8\times10^{-17}$}\\
\textcolor{black}{$0.2$} & \textcolor{black}{$2$} & \textcolor{black}{$n=0$} & \textcolor{black}{out} & \textcolor{black}{$0.496326605942-0.092389166442i$} & \textcolor{black}{$0.496326605942-0.092389166442i$} & \textcolor{black}{$2.0\times10^{-17}$}\\
\hline
\textcolor{black}{$0.7$} & \textcolor{black}{$0$} & \textcolor{black}{$j=0$} & \textcolor{black}{dec} & \textcolor{black}{$0.695445995443-0.001945341868i$} & \textcolor{black}{$0.695445995443-0.001945341868i$} & \textcolor{black}{$1.3\times10^{-17}$}\\
\textcolor{black}{$0.7$} & \textcolor{black}{$0$} & \textcolor{black}{$j=1$} & \textcolor{black}{dec} & \textcolor{black}{$0.693730467133-0.003236529475i$} & \textcolor{black}{$0.693730467133-0.003236529475i$} & \textcolor{black}{$3.3\times10^{-17}$}\\
\hline
\textcolor{black}{$0.8$} & \textcolor{black}{$0$} & \textcolor{black}{$j=0$} & \textcolor{black}{dec} & \textcolor{black}{$0.793541841410-0.003225245817i$} & \textcolor{black}{$0.793541841410-0.003225245817i$} & \textcolor{black}{$3.2\times10^{-18}$}\\
\textcolor{black}{$0.8$} & \textcolor{black}{$1$} & \textcolor{black}{$j=0$} & \textcolor{black}{dec} & \textcolor{black}{$0.782288498618-0.010162267390i$} & \textcolor{black}{$0.782288498618-0.010162267390i$} & \textcolor{black}{$3.5\times10^{-17}$} \\ [0.5ex]
\hline
\end{tabular}
\end{table*}

\textcolor{black}{The maximum ESM--CFM difference in Table~\ref{table:direct-massive-benchmark} is $1.5\times10^{-7}$ for the slowly converging $\mu=0.2$, $\ell=0$ fundamental mode. Every other displayed difference is of order $10^{-15}$ or smaller. All are well below the $10^{-4}$ acceptance tolerance used for the three-resolution spectral catalogue. The outgoing rows directly validate fundamental massive QNMs and selected overtones. The near-threshold rows show, at the frequency level, that the same complex values solve the independently imposed decaying boundary condition. Because the original spectral archive retained $\Omega$ but not its reciprocal pre-image $\Lambda$, this agreement does not retroactively identify which sheet was stored for each archived entry. For the particular large-$\mu$ stress tests and highly damped purely imaginary ladders displayed here, we have not found an equally precise pointwise external benchmark. In those sectors, our claim remains deliberately narrower, i.e. they are stable roots of the uniformised boundary-value problem under independent assembly, resolution, and precision changes. The massive comparisons above validate the differential operator, asymptotic factorisation, and low-lying spectrum in an overlap domain, but they do not by themselves establish the physical sheet, Green-function residue, or time-domain excitation of every unbenchmarked root.}

\begin{figure}[!t]
    \centering
    \includegraphics[width=0.98\linewidth]{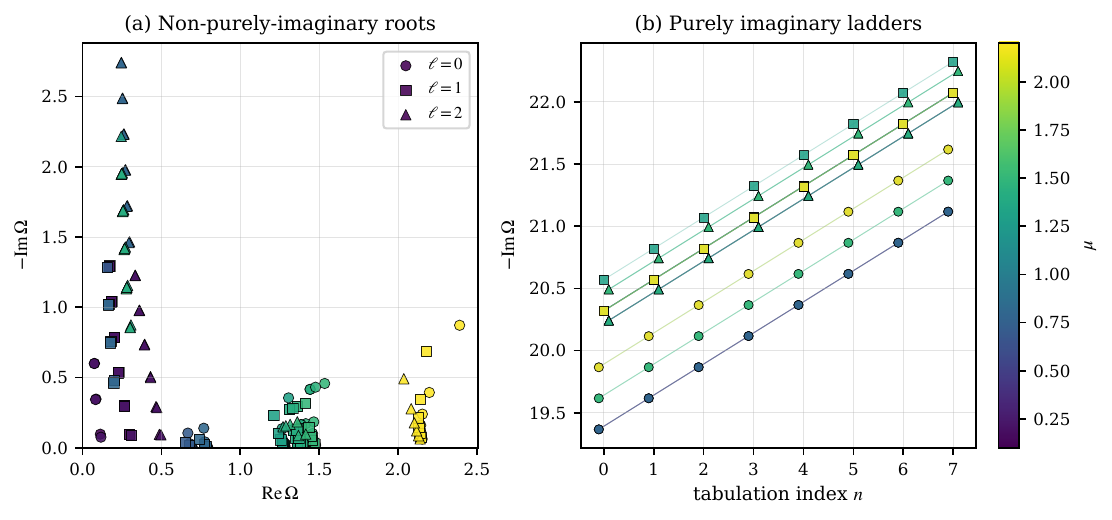}
    \caption{\textit{\textcolor{black}{Visual summary of the dimensionless couplings primarily focused upon in this work, constructed directly from the archived convergence reports. Left: stable non-purely-imaginary roots in the $(\Re {\rm e} \{\Omega\},-\Im {\rm m}\{\Omega\})$ plane for $0.1\leq\mu\leq2.2$. Right: damping rates of the purely imaginary sequences as functions of the tabulation index $n$. Marker shape distinguishes $\ell\in\{0,1,2\}$, while colour encodes $\mu$. In the right panel, thin lines connect only entries with the same $(\mu,\ell)$ and serve as guides to the eye. They do not assert a global analytic continuation between different parameter values. Exact numerical values and their ancillary diagnostics remain in Tables~\ref{tablemu0102}--\ref{tablemu138_14_142_145_150_220}.}}}
    \label{fig:spectral_catalogue_overview}
\end{figure}

\textcolor{black}{Before presenting the detailed numerical catalogue, Fig.~\ref{fig:spectral_catalogue_overview} gives a compact visual overview of the full-precision converged roots from the archived reports corresponding to the main-text spectrum tables. The figure displays the global organisation of the non-purely-imaginary roots and the near-linear damping ladders. The tables are retained because they supply the exact complex frequencies, mode indices, threshold-proxy values, spacings, unavailable entries, convergence data, and benchmark differences needed for quantitative comparison and reproducibility. No global branch tracking or Riemann-sheet assignment is inferred from the visual connections.}

\textcolor{black}{We proceed with our exploration of the parameter space in Tables~\ref{tablemu0102}, \ref{tablemu0102overdamped}, \ref{tablemu0708}, and \ref{tablemu0708overdamped}, where we list stable spectral roots for $0<\mu<1$. We use $\delta_{\rm R}=[\Re {\rm e}\{\Omega\}]^2-\mu^2$ only to organise the entries relative to the nominal real-frequency threshold. The symbols $\Omega_{>}$ and $\Omega_{<}$ denote roots with $\delta_{\rm R}>0$ and $\delta_{\rm R}<0$, respectively. As discussed in Section \ref{sec:context}, the sign of $\delta_{\rm R}$ does not itself denote propagating and evanescent physical sectors. Upon comparing Tables~\ref{tablemu0102} and \ref{tablemu0708}, the $\delta_{\rm R}<0$ group becomes increasingly prominent as the mass term raises the asymptotic potential and suppresses the photon-sphere barrier. The exact radial decay or growth of any individual root remains a $k(\Lambda)$ and sheet-dependent question.}

\begin{figure}[t]
    \centering
    \includegraphics[width=0.7\linewidth]{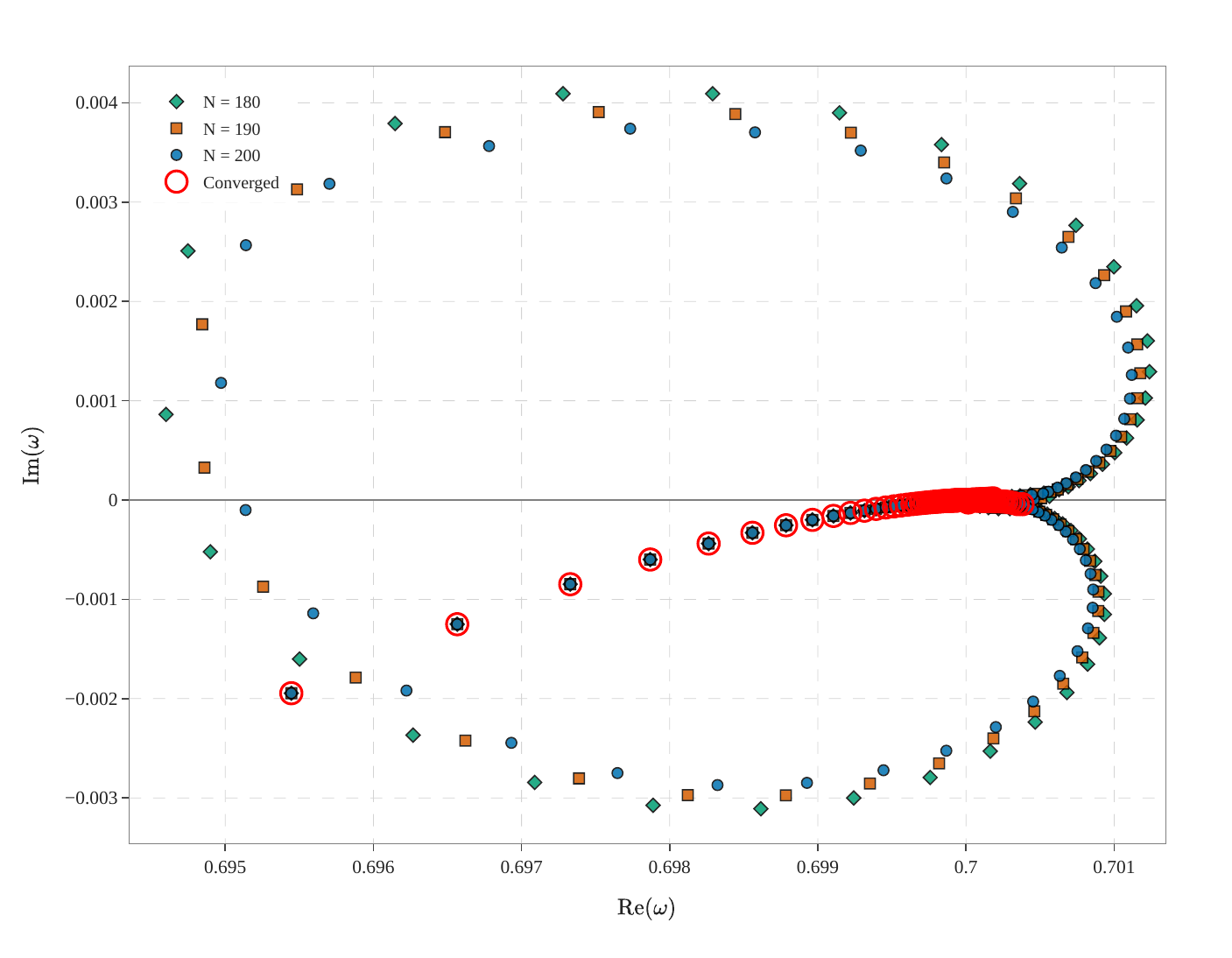}
    \caption{\textit{\textcolor{black}{Discrete stable roots with $\delta_{\rm R}<0$ for $\mu=0.7$ and $\ell=0$, computed by Chebyshev collocation with $N\in\{180,190,200\}$. The red circles indicate roots stable under the resolution test. Their clustering near $\Omega\simeq\mu$ is a numerical feature of the compactified spectral problem. The sign of $\delta_{\rm R}$ alone does not establish an evanescent, quasibound, or outgoing-resonant asymptotic class.}}}
    \label{fig:QNMplot}
\end{figure}

\textcolor{black}{In addition to damped complex roots on both sides of the nominal real-frequency threshold, we identify families of purely imaginary stable spectral roots with approximately uniform spacing in their imaginary parts. For $\mu \in\{0.1,0.2\}$ and $\mu\in\{0.7,0.8\}$, the first members are displayed in Tables~\ref{tablemu0102overdamped} and \ref{tablemu0708overdamped}, respectively. The magnitude of the imaginary component increases monotonically along each retained sequence, while the spacing is close to the Schwarzschild surface gravity scale $M\kappa_{\rm H}=1/4$. The full convergence report for the representative case $\mu=0.1$, $\ell=0$ contains $86$ matched purely imaginary roots. Table~\ref{table:imaginary-convergence} displays the individual values at all three resolutions for the first eight roots used in Table~\ref{tablemu0102overdamped}. Every maximal pairwise drift lies below the acceptance threshold $10^{-4}$, and for the more highly damped members it falls to the $10^{-6}$--$10^{-7}$ range. Their real parts are numerically zero by many tens of digits. This behaviour, repeated across the parameter reports, is inconsistent with the simplest explanation in terms of roots drifting with the spectral resolution. The near-threshold $\delta_{\rm R}<0$ cluster in Fig.~\ref{fig:QNMplot} is a separate feature and is not used as evidence for the purely imaginary ladder.}

\begin{table}[t]
\centering
\caption{\textit{\textcolor{black}{Individual-resolution convergence of the first eight purely imaginary roots for $\mu=0.1$ and $\ell=0$. The entries show the damping rate $-\Im {\rm m}\{\Omega\}$ obtained from the independently assembled calculations at $N\in\{180,190,200\}$, with the decimal working precision set equal to $N$. The last column is $\delta_n=\max_{N,N'}|\Omega_n^{(N)}-\Omega_n^{(N')}|$.}}}
\label{table:imaginary-convergence}
\setlength\tabcolsep{0.1cm}
\def\arraystretch{1.5}
\begin{tabular}{@{}|c|c|c|c|c|@{}}
\hline
& \textcolor{black}{$-\Im {\rm m}\{\Omega\}$} & \textcolor{black}{$-\Im {\rm m}\{\Omega\}$} & \textcolor{black}{$-\Im {\rm m}\{\Omega\}$} &  \\[-1ex]
\textcolor{black}{$n$} & \textcolor{black}{$N=180$} & \textcolor{black}{$N=190$} & \textcolor{black}{$N=200$} & \textcolor{black}{$\delta_n$} \\[0.5ex]
\hline
\textcolor{black}{$0$} & \textcolor{black}{$19.36673635$} & \textcolor{black}{$19.36673228$} & \textcolor{black}{$19.36676712$} & \textcolor{black}{$3.48\times10^{-5}$}\\
\textcolor{black}{$1$} & \textcolor{black}{$19.61679355$} & \textcolor{black}{$19.61680126$} & \textcolor{black}{$19.61682222$} & \textcolor{black}{$2.87\times10^{-5}$}\\
\textcolor{black}{$2$} & \textcolor{black}{$19.86684362$} & \textcolor{black}{$19.86683838$} & \textcolor{black}{$19.86680964$} & \textcolor{black}{$3.40\times10^{-5}$}\\
\textcolor{black}{$3$} & \textcolor{black}{$20.11689415$} & \textcolor{black}{$20.11689598$} & \textcolor{black}{$20.11691429$} & \textcolor{black}{$2.01\times10^{-5}$}\\
\textcolor{black}{$4$} & \textcolor{black}{$20.36694483$} & \textcolor{black}{$20.36694512$} & \textcolor{black}{$20.36693850$} & \textcolor{black}{$6.62\times10^{-6}$}\\
\textcolor{black}{$5$} & \textcolor{black}{$20.61699320$} & \textcolor{black}{$20.61699229$} & \textcolor{black}{$20.61699200$} & \textcolor{black}{$1.20\times10^{-6}$}\\
\textcolor{black}{$6$} & \textcolor{black}{$20.86704138$} & \textcolor{black}{$20.86704204$} & \textcolor{black}{$20.86704450$} & \textcolor{black}{$3.12\times10^{-6}$}\\
\textcolor{black}{$7$} & \textcolor{black}{$21.11708871$} & \textcolor{black}{$21.11708850$} & \textcolor{black}{$21.11708650$} & \textcolor{black}{$2.21\times10^{-6}$} \\ [0.5ex]
\hline
\end{tabular}
\end{table}

\textcolor{black}{Purely imaginary spectra have been reported in other black hole problems. For example, within Schwarzschild-de Sitter spacetime the spectrum splits into distinct branches, one of which is purely imaginary in the small-mass regime \cite{Konoplya2024_TwoRegimes}. Recent massive-Schwarzschild work now provides systematic Hill-determinant and continued-fraction benchmarks for the ordinary low-lying QNM branches \cite{AlvesPonquioMedeiros2025PRD}, and the comparison above incorporates those results. Neither that study nor the older recurrence analysis of Ref.~\cite{Konoplya2005PLB}, however, tabulates the highly damped imaginary-axis ladders reported here in a form permitting a pointwise external comparison. We therefore retain the explicit convergence evidence while keeping their sheet and dynamical interpretation qualified, as discussed in Section~\ref{sec:results}.}

\begin{table}
\centering
\caption{\textit{\textcolor{black}{Stable complex spectral roots for massive scalar perturbations of the Schwarzschild black hole for $\ell\in\{0,1,2\}$ and $\mu\in\{0.1,0.2\}$. The results use $200$ Chebyshev polynomials and $200$-digit precision. Here $\delta_{\rm R}=[\Re {\rm e}\{\Omega\}]^2-\mu^2$ is only a real-frequency threshold proxy. It does not determine the Riemann sheet or the asymptotic decay/growth class.}}}
\label{tablemu0102}
\setlength\tabcolsep{0.1cm}
\def\arraystretch{1.5}
\begin{tabular}{@{}|c|c|c|c|c||c|c|c|c|c|c|c|c|c|c|@{}}
\hline
\textcolor{black}{$\mu$} & \textcolor{black}{$\ell$} & \textcolor{black}{$n$} & \textcolor{black}{$\Omega$} & \textcolor{black}{$\delta_{\rm R}$} & \textcolor{black}{$\mu$} & \textcolor{black}{$\ell$} & \textcolor{black}{$n$} & \textcolor{black}{$\Omega$} & \textcolor{black}{$\delta_{\rm R}$}\\ [0.5ex]
\hline
$0.1$  &$0$    & $0$ &$0.1124-0.0968i$ &  0.0026     & $0.2$  & $0$    & $0$ & $0.1162-0.0754i$ & -0.0265\\
       &       & $1$ &$0.0849-0.3467i$ &  -0.0028     &        &        & $1$ & $0.0819-0.3427i$ & -0.0333\\
       &       & $2$ &$0.0752-0.6006i$ &  -0.0043     &        &        & $2$ & $0.0736-0.5992i$ & -0.0346\\
       \cline{2-5}
       \cline{7-10}
       &$1$    & $0$ &$0.2974-0.0950i$ &  0.0784     &        & $1$    & $0$ & $0.3110-0.0866i$ & 0.0567\\
       &       & $1$ &$0.2647-0.3029i$ &  0.0601     &        &        & $1$ & $0.2651-0.2928i$ & 0.0303\\
       &       & $2$ &$0.2287-0.5386i$ &  0.0423     &        &        & $2$ & $0.2261-0.5341i$ & 0.0111\\
       &       & $3$ &$0.2026-0.7876i$ &  0.0310     &        &        & $3$ & $0.2005-0.7857i$ & 0.0002\\
       &       & $4$ &$0.1846-1.0404i$ &  0.0241     &        &        & $4$ & $0.1832-1.0394i$ & -0.0064\\
       &       & $5$ &$0.1717-1.2939i$ &  0.0195     &        &        & $5$ & $0.1707-1.2933i$ & -0.0109\\
       \cline{2-5}
       \cline{7-10}
       &$2$    & $0$ &$0.4868-0.0957i$ & 0.2270     &        & $2$    & $0$ & $0.4963-0.0924i$ & 0.2063\\
       &       & $1$ &$0.4655-0.2933i$ &  0.2067     &        &        & $1$ & $0.4703-0.2862i$ & 0.1812\\
       &       & $2$ &$0.4307-0.5065i$ &  0.1755     &        &        & $2$ & $0.4310-0.5002i$ & 0.1458\\
       &       & $3$ &$0.3934-0.7368i$ &  0.1448     &        &        & $3$ & $0.3919-0.7328i$ & 0.1136\\
       &       & $4$ &$0.3607-0.9792i$ &  0.1201     &        &        & $4$ & $0.3591-0.9769i$ & 0.0890\\
       &       & $5$ &$0.3344-1.2279i$ &  0.1018     &        &        & $5$ & $0.3331-1.2266i$ & 0.0710\\
[1ex]
\hline 
\end{tabular}
\end{table}

\begin{table}
\centering
\caption{\textit{Purely imaginary stable spectral roots for massive scalar perturbations of the Schwarzschild black hole for $\ell\in\{0,1,2\}$ and $\mu\in\{0.1,0.2\}$. The results use $200$ polynomials and $200$-digit precision. Here, $\Delta\Omega=\Omega_n-\Omega_{n+1}$. \textcolor{black}{Algebraically, all entries have $\delta_{\rm R}=-\mu^2$, but this does not determine their Riemann sheet or whether the radial exponential decays or grows at infinity.} Entries labelled \enquote{N/A} indicate unavailable data.}}
\label{tablemu0102overdamped}
\setlength\tabcolsep{0.1cm}
\def\arraystretch{1.5}
\begin{tabular}{@{}|c|c|c|c|c||c|c|c|c|c|c|c|c|@{}}
\hline
$\mu$  &$\ell$ & $n$ &$\Omega$           & $\Delta\Omega$  & $\mu$   & $\ell$ & $n$ & $\Omega$       & $\Delta\Omega$ \\ [0.5ex]
\hline
$0.1$  &$0$    & $0$ &$0.0000-19.36677i$ & \mbox{N/A}      & $0.2$   & $0$    & $0$ & $0.0000-19.36676i$    & \mbox{N/A}\\
       &       & $1$ &$0.0000-19.61682i$ & $0.25006i$      &         &        & $1$ & $0.0000-19.61681i$    & $0.25005i$\\
       &       & $2$ &$0.0000-19.86681i$ & $0.24999i$      &         &        & $2$ & $0.0000-19.86680i$    & $0.24999i$\\
       &       & $3$ &$0.0000-20.11691i$ & $0.25010i$      &         &        & $3$ & $0.0000-20.11691i$    & $0.25010i$\\
       &       & $4$ &$0.0000-20.36694i$ & $0.25003i$      &         &        & $4$ & $0.0000-20.36693i$    & $0.25002i$\\
       &       & $5$ &$0.0000-20.61699i$ & $0.25005i$      &         &        & $5$ & $0.0000-20.61698i$    & $0.25005i$\\
       &       & $6$ &$0.0000-20.86704i$ & $0.25005i$      &         &        & $6$ & $0.0000-20.86704i$    & $0.25005i$\\
       &       & $7$ &$0.0000-21.11709i$ & $0.25004i$      &         &        & $7$ & $0.0000-21.11708i$    & $0.25004i$\\
       \cline{2-5}
       \cline{7-10}
       &$1$    & $0$ &$0.0000-20.32006i$ & $0.25028i$      &         & $1$    & $0$ & $0.0000-20.32005i$    & $0.25028i$\\
       &       & $1$ &$0.0000-20.57037i$ & $0.25031i$      &         &        & $1$ & $0.0000-20.57036i$    & $0.25031i$\\
       &       & $2$ &$0.0000-20.82070i$ & $0.25033i$      &         &        & $2$ & $0.0000-20.82070i$    & $0.25033i$\\
       &       & $3$ &$0.0000-21.07099i$ & $0.25029i$      &         &        & $3$ & $0.0000-21.07098i$    & $0.25029i$\\
       &       & $4$ &$0.0000-21.32130i$ & $0.25031i$      &         &        & $4$ & $0.0000-21.32129i$    & $0.25031i$\\
       &       & $5$ &$0.0000-21.57160i$ & $0.25030i$      &         &        & $5$ & $0.0000-21.57159i$    & $0.25030i$\\
       &       & $6$ &$0.0000-21.82189i$ & $0.25029i$      &         &        & $6$ & $0.0000-21.82188i$    & $0.25029i$\\
       &       & $7$ &$0.0000-22.07218i$ & $0.25029i$      &         &        & $7$ & $0.0000-22.07217i$    & $0.25029i$\\
       \cline{2-5}
       \cline{7-10}
       &$2$    & $0$ &$0.0000-20.24222i$ & $0.25071i$      &         & $2$    & $0$ & $0.0000-20.24221i$    & $0.25071i$\\
       &       & $1$ &$0.0000-20.49275i$ & $0.25053i$      &         &        & $1$ & $0.0000-20.49274i$    & $0.25053i$\\
       &       & $2$ &$0.0000-20.74348i$ & $0.25073i$      &         &        & $2$ & $0.0000-20.74347i$    & $0.25073i$\\
       &       & $3$ &$0.0000-20.99408i$ & $0.25060i$      &         &        & $3$ & $0.0000-20.99407i$    & $0.25060i$\\
       &       & $4$ &$0.0000-21.24470i$ & $0.25062i$      &         &        & $4$ & $0.0000-21.24469i$    & $0.25062i$\\
       &       & $5$ &$0.0000-21.49533i$ & $0.25062i$      &         &        & $5$ & $0.0000-21.49531i$    & $0.25063i$\\
       &       & $6$ &$0.0000-21.74592i$ & $0.25059i$      &         &        & $6$ & $0.0000-21.74591i$    & $0.25059i$\\
       &       & $7$ &$0.0000-21.99652i$ & $0.25060i$      &         &        & $7$ & $0.0000-21.99651i$    & $0.25060i$\\
[1ex]
\hline 
\end{tabular}
\end{table}

\begin{table}
\centering
\caption{\textit{\textcolor{black}{Stable complex spectral roots for massive scalar perturbations of the Schwarzschild black hole for $\ell\in\{0,1,2\}$ and $\mu\in\{0.7,0.8\}$. The columns $\Omega_{>}$ and $\Omega_{<}$ denote $\delta_{\rm R}>0$ and $\delta_{\rm R}<0$, respectively, where $\delta_{\rm R}=[\Re {\rm e}\{\Omega\}]^2-\mu^2$. Entries labelled \enquote{N/A} indicate unavailable data.}}}
\label{tablemu0708}
\setlength\tabcolsep{0.1cm}
\def\arraystretch{1.5}
\begin{tabular}{@{}|c|c|c|c|c|c||c|c|c|c|c|c|c|c|c|@{}}
\hline
\textcolor{black}{$\mu$} & \textcolor{black}{$\ell$} & \textcolor{black}{$n$} & \textcolor{black}{$\Omega_{>}$} & \textcolor{black}{$\Omega_{<}$} & \textcolor{black}{$\delta_{\rm R}$} & \textcolor{black}{$\mu$} & \textcolor{black}{$\ell$} & \textcolor{black}{$n$} & \textcolor{black}{$\Omega_{>}$} & \textcolor{black}{$\Omega_{<}$} & \textcolor{black}{$\delta_{\rm R}$}\\ [0.5ex]
\hline
$0.7$  &$0$    & $0$ &\mbox{N/A}   &$0.6954-0.0019i$ & \textcolor{black}{$-0.0064$}  & $0.8$  & $0$    & $0$ & \mbox{N/A} &$0.7935-0.0032i$ & \textcolor{black}{$-0.0104$}\\
       &       & $1$ &\mbox{N/A}   &$0.6937-0.0032i$ & \textcolor{black}{$-0.0088$}  &        &        & $1$ & \mbox{N/A} &$0.7913-0.0053i$ & \textcolor{black}{$-0.0138$}\\
       &       & $2$ &\mbox{N/A}   &\textcolor{black}{$0.6910-0.0059i$} & \textcolor{black}{$-0.0125$} &        &        & $2$ & \mbox{N/A} &$0.7879-0.0093i$ & \textcolor{black}{$-0.0192$}\\
       &       & $3$ &\mbox{N/A}   &$0.6864-0.0120i$ & \textcolor{black}{$-0.0189$}  &        &        & $3$ & \mbox{N/A} &$0.7824-0.0184i$ & \textcolor{black}{$-0.0279$}\\
       &       & $4$ &\mbox{N/A}   &$0.6787-0.0295i$ & \textcolor{black}{$-0.0294$}  &        &        & $4$ & \mbox{N/A} &$0.7745-0.0424i$ & \textcolor{black}{$-0.0401$}\\
       &       & $5$ &\mbox{N/A}   &$0.6673-0.1046i$ & \textcolor{black}{$-0.0447$}  &        &        & $5$ & \mbox{N/A} &$0.7682-0.1392i$ & \textcolor{black}{$-0.0499$}\\
        \cline{2-6}
        \cline{8-12}
 &$1$    & $0$ &\mbox{N/A}   &\textcolor{black}{$0.6775-0.0129i$} & \textcolor{black}{$-0.0310$}  &        & $1$    & $0$ & \mbox{N/A} &\textcolor{black}{$0.7823-0.0102i$} & \textcolor{black}{$-0.0280$}\\
       &       & $1$ &\mbox{N/A}   &$0.6508-0.0361i$ & \textcolor{black}{$-0.0665$} &        &        & $1$ & \mbox{N/A} &$0.7704-0.0222i$ & \textcolor{black}{$-0.0465$}\\
       &       & $2$ &\mbox{N/A}   &$0.2011-0.4800i$ & \textcolor{black}{$-0.4496$}  &        &        & $2$ & \mbox{N/A} &$0.7393-0.0595i$ & \textcolor{black}{$-0.0934$} \\
       &       & $3$ &\mbox{N/A}   &$0.1792-0.7571i$ & \textcolor{black}{$-0.4579$} &        &        & $3$ & \mbox{N/A} &$0.1970-0.4634i$ & \textcolor{black}{$-0.6012$}\\
       &       & $4$ &\mbox{N/A}   &$0.1669-1.0225i$ & \textcolor{black}{$-0.4621$}  &        &        & $4$ & \mbox{N/A} &$0.1747-0.7475i$ & \textcolor{black}{$-0.6095$}\\
       &       & $5$ &\mbox{N/A}   &$0.1582-1.2823i$ & \textcolor{black}{$-0.4650$}  &        &        & $5$ & \mbox{N/A} &$0.1629-1.0163i$ & \textcolor{black}{$-0.6135$}\\
        \cline{2-6}
        \cline{8-12}
&$2$    & $0$ &\mbox{N/A}   &$0.2978-1.4655i$ & \textcolor{black}{$-0.4013$}  &        & $2$    & $0$ & \mbox{N/A} &$0.2937-1.4610i$ & \textcolor{black}{$-0.5537$}\\
       &       & $1$ &\mbox{N/A}   &$0.2839-1.7225i$ & \textcolor{black}{$-0.4094$}  &        &        & $1$ & \mbox{N/A} &$0.2804-1.7192i$ & \textcolor{black}{$-0.5614$}\\
       &       & $2$ &\mbox{N/A}   &$0.2725-1.9784i$ & \textcolor{black}{$-0.4157$} &        &        & $2$ & \mbox{N/A} &$0.2694-1.9759i$ & \textcolor{black}{$-0.5674$}\\
       &       & $3$ &\mbox{N/A}   &$0.2625-2.2334i$ & \textcolor{black}{$-0.4211$}  &        &        & $3$ & \mbox{N/A} &$0.2600-2.2315i$ & \textcolor{black}{$-0.5724$}\\
       &       & $4$ &\mbox{N/A}   &$0.2540-2.4879i$ & \textcolor{black}{$-0.4255$}  &        &        & $4$ & \mbox{N/A} &$0.2519-2.4863i$ & \textcolor{black}{$-0.5765$}\\
       &       & $5$ &\mbox{N/A}   &$0.2466-2.7418i$ & \textcolor{black}{$-0.4292$} &        &        & $5$ & \mbox{N/A} &$0.2447-2.7405i$ & \textcolor{black}{$-0.5801$}\\
[1ex]
\hline 
\end{tabular}
\end{table}

\begin{table}
\centering
\caption{\textit{\textcolor{black}{Purely imaginary stable spectral roots for massive scalar perturbations of the Schwarzschild black hole for $\ell\in\{0,1,2\}$ and $\mu\in\{0.7,0.8\}$. The results use $200$ polynomials and $200$-digit precision, and $\Delta\Omega=\Omega_n-\Omega_{n+1}$. All entries have $\delta_{\rm R}=-\mu^2$, but this algebraic identity does not determine their sheet or exponential behaviour at infinity. Entries labelled \enquote{N/A} indicate unavailable data.}}}
\label{tablemu0708overdamped}
\setlength\tabcolsep{0.1cm}
\def\arraystretch{1.5}
\begin{tabular}{@{}|c|c|c|c|c||c|c|c|c|c|c|c|c|@{}}
\hline
$\mu$  &$\ell$ & $n$ &$\Omega$            & $\Delta\Omega$  & $\mu$   & $\ell$ & $n$ & $\Omega$       & $\Delta\Omega$ \\ [0.5ex]
\hline
$0.7$  &$0$    & $0$ &$0.0000-19.36665i$  & \mbox{N/A}       & $0.8$   & $0$    & $0$ & $0.0000-19.36661i$    & \mbox{N/A}\\
       &       & $1$ &$0.0000-19.61669i$  & $0.25005i$      &         &        & $1$ & $0.0000-19.61665i$    & $0.25004i$\\
       &       & $2$ &$0.0000-19.86669i$  & $0.24999i$      &         &        & $2$ & $0.0000-19.86665i$    & $0.25000i$\\
       &       & $3$ &$0.0000-20.11679i$  & $0.25011i$      &         &        & $3$ & $0.0000-20.11675i$    & $0.25011i$\\
       &       & $4$ &$0.0000-20.36682i$  & $0.25002i$      &         &        & $4$ & $0.0000-20.36678i$    & $0.25002i$\\
       &       & $5$ &$0.0000-20.61687i$  & $0.25006i$      &         &        & $5$ & $0.0000-20.61684i$    & $0.25006i$\\
       &       & $6$ &$0.0000-20.86693i$  & $0.25005i$      &         &        & $6$ & $0.0000-20.86689i$    & $0.25005i$\\
       &       & $7$ &$0.0000-21.11697i$  & $0.25004i$      &         &        & $7$ & $0.0000-21.11694i$    & $0.25004i$\\
       \cline{2-5}
       \cline{7-10}
       &$1$    & $0$ &$0.0000-20.31992i$  & $0.25027i$      &         & $1$    & $0$ & $0.0000-20.31988i$    & $0.25027i$\\
       &       & $1$ &$0.0000-20.57023i$  & $0.25032i$      &         &        & $1$ & $0.0000-20.57019i$    & $0.25032i$\\
       &       & $2$ &$0.0000-20.82057i$  & $0.25034i$      &         &        & $2$ & $0.0000-20.82053i$    & $0.25034i$\\
       &       & $3$ &$0.0000-21.07086i$  & $0.25029i$      &         &        & $3$ & $0.0000-21.07082i$    & $0.25030i$\\
       &       & $4$ &$0.0000-21.32118i$  & $0.25031i$      &         &        & $4$ & $0.0000-21.32114i$    & $0.25031i$\\
       &       & $5$ &$0.0000-21.57147i$  & $0.25030i$      &         &        & $5$ & $0.0000-21.57144i$    & $0.25030i$\\
       &       & $6$ &$0.0000-21.82177i$  & $0.25029i$      &         &        & $6$ & $0.0000-21.82173i$    & $0.25029i$\\
       &       & $7$ &$0.0000-22.07206i$  & $0.25029i$      &         &        & $7$ & $0.0000-22.07202i$    & $0.25029i$\\
       \cline{2-5}
       \cline{7-10}
       &$2$    & $0$ &$0.0000-20.24202i$  & $0.25069i$      &         & $2$    & $0$ & $0.0000-20.24196i$    & $0.25069i$\\
       &       & $1$ &$0.0000-20.49256i$  & $0.25053i$      &         &        & $1$ & $0.0000-20.49249i$    & $0.25053i$\\
       &       & $2$ &$0.0000-20.74330i$  & $0.25074i$      &         &        & $2$ & $0.0000-20.74324i$    & $0.25074i$\\
       &       & $3$ &$0.0000-20.99389i$  & $0.25060i$      &         &        & $3$ & $0.0000-20.99383i$    & $0.25060i$\\
       &       & $4$ &$0.0000-21.24452i$  & $0.25063i$      &         &        & $4$ & $0.0000-21.24446i$    & $0.25063i$\\
       &       & $5$ &$0.0000-21.49515i$  & $0.25063i$      &         &        & $5$ & $0.0000-21.49509i$    & $0.25063i$\\
       &       & $6$ &$0.0000-21.74574i$  & $0.25060i$      &         &        & $6$ & $0.0000-21.74569i$    & $0.25060i$\\
       &       & $7$ &$0.0000-21.99635i$  & $0.25060i$      &         &        & $7$ & $0.0000-21.99629i$    & $0.25060i$\\
[1ex]
\hline 
\end{tabular}
\end{table}

\begin{table}[h]
\centering
\caption{\textit{\textcolor{black}{Stable complex spectral roots with $\delta_{\rm R}<0$, denoted $\Omega_{<}$, for $\mu\in\{1.45,1.50,2.20\}$ and $\ell\in\{0,1,2\}$. The quantity $\delta_{\rm R}=[\Re {\rm e}\{\Omega\}]^2-\mu^2$ is a neutral real-frequency threshold proxy only. No outgoing, decaying, propagating, evanescent, or quasibound assignment is inferred from its sign.}}}
\label{tablemu145_150_220-straight}
\vspace*{1em}
\setlength\tabcolsep{0.1cm}
\def\arraystretch{1.5}
\begin{tabular}{|c|c|c|c|c||c|c|c|c|c||c|c|c|c|c|}
\hline
\textcolor{black}{$\mu$} & \textcolor{black}{$\ell$} & \textcolor{black}{$n$} & \textcolor{black}{$\Omega_{<}$} & \textcolor{black}{$\delta_{\rm R}$} & \textcolor{black}{$\mu$} & \textcolor{black}{$\ell$} & \textcolor{black}{$n$} & \textcolor{black}{$\Omega_{<}$} & \textcolor{black}{$\delta_{\rm R}$} & \textcolor{black}{$\mu$} & \textcolor{black}{$\ell$} & \textcolor{black}{$n$} & \textcolor{black}{$\Omega_{<}$} & \textcolor{black}{$\delta_{\rm R}$}\\[0.5ex]
\hline
\textcolor{black}{$1.45$} & 0 & 0 & 1.4289-0.0174i & -0.0607 & \textcolor{black}{$1.50$} & 0 & 0 & 1.4774-0.0194i & -0.0673 & \textcolor{black}{$2.20$} & 0 & 0 & 2.1531-0.0620i & -0.2042 \\
     &   & 1 & 1.4246-0.0247i & -0.0730  &     &   & 1 & 1.4729-0.0274i & -0.0806 &     &   & 1 & 2.1481-0.0826i & -0.2257 \\
     &   & 2 & 1.4195-0.0365i & -0.0875 &     &   & 2 & 1.4676-0.0403i & -0.0962 &     &   & 2 & 2.1441-0.1131i & -0.2428 \\
     &   & 3 & 1.4138-0.0564i & -0.1037 &     &   & 3 & 1.4620-0.0618i & -0.1126 &     &   & 3 & 2.1439-0.1603i & -0.2437 \\
     &   & 4 & 1.4095-0.0931i & -0.1158 &     &   & 4 & 1.4585-0.1012i & -0.1228 &     &   & 4 & 2.1542-0.2393i & -0.1994 \\
     &   & 5 & 1.4151-0.1716i & -0.1000    &     &   & 5 & 1.4659-0.1843i & -0.1011 &     &   & 5 & 2.1984-0.3935i & -0.0070  \\
    \cline{2-5}
    \cline{7-10}
    \cline{12-15}
     & 1 & 0 & 1.4242-0.0203i & -0.0742 &     & 1 & 0 & 1.4723-0.0228i & -0.0823 &     & 1 & 0 & 2.1446-0.0746i & -0.2407 \\
     &   & 1 & 1.4180-0.0298i & -0.0918 &     &   & 1 & 1.4659-0.0333i & -0.1011 &     &   & 1 & 2.1378-0.1020i & -0.2698 \\
     &   & 2 & 1.4099-0.0459i & -0.1147 &     &   & 2 & 1.4577-0.0508i & -0.1251 &     &   & 2 & 2.1319-0.1442i & -0.2950  \\
     &   & 3 & 1.3995-0.0750i & -0.1439 &     &   & 3 & 1.4474-0.0823i & -0.1550  &     &   & 3 & 2.1299-0.2139i & -0.3035 \\
     &   & 4 & 1.3865-0.1349i & -0.1801 &     &   & 4 & 1.4352-0.1466i & -0.1902 &     &   & 4 & 2.1403-0.3456i & -0.2591 \\
     &   & 5 & 1.3627-0.2952i & -0.2455 &     &   & 5 & 1.4140-0.3179i & -0.2506 &     &   & 5 & 2.1778-0.6852i & -0.0972 \\
    \cline{2-5}
    \cline{7-10}
    \cline{12-15}
     & 2 & 0 & 1.3685-0.0874i & -0.2297 &     & 2 & 0 & 1.4148-0.0974i & -0.2483 &     & 2 & 0 & 2.1405-0.0631i & -0.2583 \\
     &   & 1 & 1.3159-0.1670i & -0.3709 &     &   & 1 & 1.3613-0.1849i & -0.3969 &     &   & 1 & 2.1303-0.0859i & -0.3018 \\
     &   & 2 & 0.2789-1.1387i & -2.0247 &     &   & 2 & 0.2768-1.1323i & -2.1734 &     &   & 2 & 2.1181-0.1207i & -0.3537 \\
     &   & 3 & 0.2649-1.4153i & -2.0323 &     &   & 3 & 0.2629-1.4104i & -2.1809 &     &   & 3 & 2.1033-0.1771i & -0.4161 \\
     &   & 4 & 0.2547-1.6844i & -2.0376 &     &   & 4 & 0.2529-1.6807i & -2.1860  &     &   & 4 & 2.0833-0.2778i & -0.4999 \\
     &   & 5 & 0.2467-1.9488i & -2.0416 &     &   & 5 & 0.2450-1.9457i & -2.1900   &     &   & 5 & 2.0365-0.4915i & -0.6927\\[1ex]
\hline 
\end{tabular}
\end{table}

\begin{table}
\centering
\caption{\textit{\textcolor{black}{Purely imaginary stable spectral roots for $\mu\in\{1.30,1.40,1.42\}$ and $\ell\in\{0,1,2\}$. The results use $200$ polynomials and $200$-digit precision, with $\Delta\Omega=\Omega_n-\Omega_{n+1}$. All entries have $\delta_{\rm R}=-\mu^2$, but this does not determine the Riemann sheet or the sign of $\Im {\rm m}\{k\}$. Entries labelled \enquote{N/A} indicate unavailable data.}}}
\label{tablemu138_14_142overdamped}
\vspace*{1em}
\def\arraystretch{1.5}
\begin{tabular}{@{}|c|c|c|c|c||c|c|c|c|c||c|c|c|c|c|@{}}
\hline
$\mu$  &$\ell$ & $n$ &$\Omega$         & $\Delta\Omega$  & $\mu$   & $\ell$ & $n$ & $\Omega$       & $\Delta\Omega$ & $\mu$   & $\ell$ & $n$ & $\Omega$ & $\Delta\Omega$ \\ [0.5ex]
\hline
$1.30$ &$0$ & $0$ &$0.0000-19.86636i$& $0.25001i$ & $1.40$ & $0$ & $0$ & \mbox{N/A}        & \mbox{N/A} & $1.42$  & $0$& $0$ & \mbox{N/A}    & \mbox{N/A}\\
       &    & $1$ &$0.0000-20.11647i$& $0.25012i$ &        &     & $1$ & \mbox{N/A}        & \mbox{N/A} &         &    & $1$ & \mbox{N/A}    & \mbox{N/A}\\
       &    & $2$ &$0.0000-20.36650i$& $0.25003i$ &        &     & $2$ & \mbox{N/A}        & \mbox{N/A} &         &    & $2$ & \mbox{N/A}    & \mbox{N/A}\\
       &    & $3$ &$0.0000-20.61657i$& $0.25007i$ &        &     & $3$ & \mbox{N/A}        & \mbox{N/A} &         &    & $3$ & \mbox{N/A}    & \mbox{N/A}\\
       &    & $4$ &$0.0000-20.86663i$& $0.25006i$ &        &     & $4$ & \mbox{N/A}        & \mbox{N/A} &         &    & $4$ & \mbox{N/A}    & \mbox{N/A}\\
       &    & $5$ &$0.0000-21.11668i$& $0.25005i$ &        &     & $5$ & \mbox{N/A}        & \mbox{N/A} &         &    & $5$ & \mbox{N/A}    & \mbox{N/A}\\
       &    & $6$ &$0.0000-21.36673i$& $0.25006i$ &        &     & $6$ & \mbox{N/A}        & \mbox{N/A} &         &    & $6$ & \mbox{N/A}    & \mbox{N/A}\\
       &    & $7$ &$0.0000-21.61678i$& $0.25005i$ &        &     & $7$ & \mbox{N/A}        & \mbox{N/A} &         &    & $7$ & \mbox{N/A}    & \mbox{N/A}\\
       \cline{2-5}
       \cline{7-10}
       \cline{12-15}
       &$1$ & $0$ &$0.0000-20.56989i$& $0.25033i$ &        & $1$ & $0$ & $0.0000-20.31947i$& $0.25027i$ &         &    & $0$ & $0.0000-20.31946i$& $0.25023i$\\
       &    & $1$ &$0.0000-20.82023i$& $0.25034i$ &        &     & $1$ & $0.0000-20.56981i$& $0.25033i$ &         &    & $1$ & $0.0000-20.56979i$& $0.25033i$\\
       &    & $2$ &$0.0000-21.07053i$& $0.25030i$ &        &     & $2$ & $0.0000-20.82015i$& $0.25034i$ &         &    & $2$ & $0.0000-20.82013i$& $0.25034i$\\
       &    & $3$ &$0.0000-21.32085i$& $0.25032i$ &        &     & $3$ & $0.0000-21.07045i$& $0.25030i$ &         &    & $3$ & $0.0000-21.07044i$& $0.25030i$\\
       &    & $4$ &$0.0000-21.57115i$& $0.25031i$ &        &     & $4$ & $0.0000-21.32077i$& $0.25032i$ &         &    & $4$ & $0.0000-21.32076i$& $0.25032i$\\
       &    & $5$ &$0.0000-21.82145i$& $0.25030i$ &        &     & $5$ & $0.0000-21.57108i$& $0.25031i$ &         &    & $5$ & $0.0000-21.57106i$& $0.25031i$\\
       &    & $6$ &$0.0000-22.07175i$& $0.25030i$ &        &     & $6$ & $0.0000-21.82138i$& $0.25030i$ &         &    & $6$ & $0.0000-21.82136i$& $0.25030i$\\
       &    & $7$ &$0.0000-22.32204i$& $0.25029i$ &        &     & $7$ & $0.0000-22.07168i$& $0.25030i$ &         &    & $7$ & $0.0000-22.07166i$& $0.25030i$\\
       \cline{2-5}
       \cline{7-10}
       \cline{12-15}
       &$2$ & $0$ &$0.0000-20.49205i$& $0.25055i$ &        & $2$ & $0$ & $0.0000-20.24138i$& $0.25065i$ &         &    & $0$ & $0.0000-20.24135i$& $0.25064i$\\
       &    & $1$ &$0.0000-20.74281i$& $0.25076i$ &        &     & $1$ & $0.0000-20.49194i$& $0.25056i$ &         &    & $1$ & $0.0000-20.49191i$& $0.25056i$\\
       &    & $2$ &$0.0000-20.99341i$& $0.25060i$ &        &     & $2$ & $0.0000-20.74270i$& $0.25076i$ &         &    & $2$ & $0.0000-20.74267i$& $0.25076i$\\
       &    & $3$ &$0.0000-21.24405i$& $0.25064i$ &        &     & $3$ & $0.0000-20.99329i$& $0.25060i$ &         &    & $3$ & $0.0000-20.99327i$& $0.25060i$\\
       &    & $4$ &$0.0000-21.49468i$& $0.25064i$ &        &     & $4$ & $0.0000-21.24394i$& $0.25064i$ &         &    & $4$ & $0.0000-21.24392i$& $0.25065i$\\
       &    & $5$ &$0.0000-21.74529i$& $0.25061i$ &        &     & $5$ & $0.0000-21.49458i$& $0.25064i$ &         &    & $5$ & $0.0000-21.49455i$& $0.25064i$\\
       &    & $6$ &$0.0000-21.99590i$& $0.25061i$ &        &     & $6$ & $0.0000-21.74518i$& $0.25061i$ &         &    & $6$ & $0.0000-21.74516i$& $0.25061i$\\
       &    & $7$ &$0.0000-22.24650i$& $0.25060i$ &        &     & $7$ & $0.0000-21.99580i$& $0.25061i$ &         &    & $7$ & $0.0000-21.99578i$& $0.25061i$\\
[1ex]
\hline 
\end{tabular}
\end{table}

\begin{table}[h]
\centering
\caption{\textit{\textcolor{black}{Stable complex spectral roots with $\delta_{\rm R}<0$, denoted $\Omega_{<}$, for $\mu\in\{1.30,1.40,1.42\}$ and $\ell\in\{0,1,2\}$. The quantity $\delta_{\rm R}=[\Re {\rm e}\{\Omega\}]^2-\mu^2$ is a neutral real-frequency threshold proxy only. No outgoing, decaying, propagating, evanescent, or quasibound assignment is inferred from its sign.}}}
\label{tablemu138_14_142-straight}
\setlength\tabcolsep{0.1cm}
\def\arraystretch{1.5}
\vspace*{1em}
\begin{tabular}{|c|c|c|c|c||c|c|c|c|c||c|c|c|c|c|}
\hline
\textcolor{black}{$\mu$} & \textcolor{black}{$\ell$} & \textcolor{black}{$n$} & \textcolor{black}{$\Omega_{<}$} & \textcolor{black}{$\delta_{\rm R}$} & \textcolor{black}{$\mu$} & \textcolor{black}{$\ell$} & \textcolor{black}{$n$} & \textcolor{black}{$\Omega_{<}$} & \textcolor{black}{$\delta_{\rm R}$} & \textcolor{black}{$\mu$} & \textcolor{black}{$\ell$} & \textcolor{black}{$n$} & \textcolor{black}{$\Omega_{<}$} & \textcolor{black}{$\delta_{\rm R}$}\\[0.5ex]
\hline
\textcolor{black}{$1.30$} & 0 & 0 & 1.2835-0.0121i & -0.0426 & \textcolor{black}{$1.40$} & 0 & 0 & 1.3999-0.0162i & -0.0003 & \textcolor{black}{$1.42$} & 0 & 0 & 1.3999-0.0162i & -0.0567 \\
    &   & 1 & \textcolor{black}{$1.2797-0.0175i$} & -0.0524 &     &   & 1 & 1.3957-0.0233i & -0.012  &      &   & 1 & 1.3957-0.0232i & -0.0684 \\
    &   & 2 & 1.2750-0.0263i & -0.0644 &     &   & 2 & 1.3906-0.0343i & -0.0262 &      &   & 2 & 1.3906-0.0343i & -0.0826 \\
    &   & 3 & 1.2693-0.0418i & -0.0789 &     &   & 3 & 1.3849-0.0533i & -0.0421 &      &   & 3 & 1.3849-0.0533i & -0.0985 \\
    &   & 4 & 1.2637-0.0710i & -0.0931 &     &   & 4 & 1.3803-0.0885i & -0.0548 &      &   & 4 & 1.3803-0.0885i & -0.1112 \\
    &   & 5 & 1.2643-0.1357i & -0.0915 &     &   & 5 & 1.3848-0.1642i & -0.0423 &      &   & 5 & 1.3848-0.1642i & -0.0987 \\
    \cline{2-5}
    \cline{7-10}
    \cline{12-15}
    & 1 & 0 & 1.2797-0.0138i & -0.0524 &     & 1 & 0 & 1.3760-0.0180i & -0.0666 &      & 1 & 0 & 1.3953-0.0189i & -0.0695 \\
    &   & 1 & 1.2742-0.0207i & -0.0664 &     &   & 1 & 1.3701-0.0266i & -0.0828 &      &   & 1 & 1.3892-0.0279i & -0.0865 \\
    &   & 2 & 1.2667-0.0326i & -0.0855 &     &   & 2 & 1.3622-0.0412i & -0.1044 &      &   & 2 & 1.3812-0.0430i & -0.1087 \\
    &   & 3 & 1.2563-0.0548i & -0.1117 &     &   & 3 & 1.3517-0.0680i & -0.1329 &      &   & 3 & 1.3708-0.0707i & -0.1373 \\
    &   & 4 & 1.2415-0.1020i & -0.1487 &     &   & 4 & 1.3380-0.1235i & -0.1698 &      &   & 4 & 1.3573-0.1280i & -0.1741 \\
    &   & 5 & 1.2115-0.2305i & -0.2223 &     &   & 5 & 1.3118-0.2731i & -0.2392 &      &   & 5 & 1.3321-0.2819i & -0.2419 \\
    \cline{2-5}
    \cline{7-10}
    \cline{12-15}
    & 2 & 0 & 0.3064-0.8742i & -1.5961 &     & 2 & 0 & 1.2709-0.1497i & -0.3448 &      & 2 & 0 & 1.2888-0.1566i & -0.3554 \\
    &   & 1 & 0.2854-1.1564i & -1.6085 &     &   & 1 & 0.3022-0.8584i & -1.8687 &      &   & 1 & 0.3014-0.8552i & -1.9256 \\
    &   & 2 & 0.2713-1.4287i & -1.6164 &     &   & 2 & 0.2810-1.1449i & -1.881  &      &   & 2 & 0.2803-1.1424i & -1.9378 \\
    &   & 3 & 0.2607-1.6948i & -1.6220  &     &   & 3 & 0.2670-1.4199i & -1.8887 &      &   & 3 & 0.2662-1.4181i & -1.9455 \\
    &   & 4 & 0.2522-1.9569i & -1.6264 &     &   & 4 & 0.2567-1.6880i & -1.8941 &      &   & 4 & 0.2559-1.6866i & -1.9509 \\
    &   & 5 & 0.2450-2.2164i & -1.6300   &     &   & 5 & 0.2485-1.9516i & -1.8982 &      &   & 5 & 0.2478-1.9505i & -1.9550 \\[1ex]
\hline
\end{tabular}
\end{table}

\begin{table}
\centering
\caption{\textit{\textcolor{black}{Purely imaginary stable spectral roots for $\mu\in\{1.45,1.50,2.20\}$ and $\ell\in\{0,1,2\}$. The results use $200$ polynomials and $200$-digit precision, with $\Delta\Omega=\Omega_n-\Omega_{n+1}$. All entries have $\delta_{\rm R}=-\mu^2$, but this does not determine the Riemann sheet or the sign of $\Im {\rm m}\{k\}$. Entries labelled \enquote{N/A} indicate unavailable data.}}}
\label{tablemu145_150_220overdamped}
\vspace*{1em}
\def\arraystretch{1.5}
\begin{tabular}{@{}|c|c|c|c|c||c|c|c|c|c||c|c|c|c|c|@{}}
\hline
$\mu$  &$\ell$ & $n$ &$\Omega$         & $\Delta\Omega$  & $\mu$   & $\ell$ & $n$ & $\Omega$       & $\Delta\Omega$ & $\mu$   & $\ell$ & $n$ & $\Omega$ & $\Delta\Omega$ \\ [0.5ex]
\hline
$1.45$ &$0$& $0$ &$0.0000-19.61622i$&$0.25001i$& $1.50$& $0$& $0$ & $0.0000-19.61618i$&$0.25000i$&$2.20$&$0$& $0$ & $0.0000-19.86540i$&$0.25008i$\\
       &   & $1$ &$0.0000-19.86624i$&$0.25002i$&       &    & $1$ & $0.0000-19.86620i$&$0.25002i$&      &   & $1$ & $0.0000-20.11553i$&$0.25013i$\\
       &   & $2$ &$0.0000-20.11636i$&$0.25012i$&       &    & $2$ & $0.0000-20.11632i$&$0.25012i$&      &   & $2$ & $0.0000-20.36557i$&$0.25004i$\\
       &   & $3$ &$0.0000-20.36639i$&$0.25003i$&       &    & $3$ & $0.0000-20.36635i$&$0.25003i$&      &   & $3$ & $0.0000-20.61566i$&$0.25009i$\\
       &   & $4$ &$0.0000-20.61646i$&$0.25007i$&       &    & $4$ & $0.0000-20.61642i$&$0.25007i$&      &   & $4$ & $0.0000-20.86574i$&$0.25007i$\\
       &   & $5$ &$0.0000-20.86653i$&$0.25006i$&       &    & $5$ & $0.0000-20.86648i$&$0.25006i$&      &   & $5$ & $0.0000-21.11581i$&$0.25007i$\\
       &   & $6$ &$0.0000-21.11657i$&$0.25005i$&       &    & $6$ & $0.0000-21.11653i$&$0.25005i$&      &   & $6$ & $0.0000-21.36588i$&$0.25008i$\\
       &   & $7$ &$0.0000-21.36663i$&$0.25006i$&       &    & $7$ & $0.0000-21.36659i$&$0.25006i$&      &   & $7$ & $0.0000-21.61595i$&$0.25007i$\\
       \cline{2-5}
       \cline{7-10}
       \cline{12-15}
       &$1$& $0$ &$0.0000-20.31943i$&$0.25027i$&       & $1$& $0$ & $0.0000-20.31938i$&$0.25026i$&      & $1$  & $0$ & $0.0000-20.31850i$&$0.25025i$\\
       &   & $1$ &$0.0000-20.56977i$&$0.25034i$&       &    & $1$ & $0.0000-20.56972i$&$0.25034i$&      &   & $1$ & $0.0000-20.56887i$&$0.25038i$\\
       &   & $2$ &$0.0000-20.82011i$&$0.25034i$&       &    & $2$ & $0.0000-20.82006i$&$0.25034i$&      &   & $2$ & $0.0000-20.81923i$&$0.25036i$\\
       &   & $3$ &$0.0000-21.07041i$&$0.25030i$&       &    & $3$ & $0.0000-21.07037i$&$0.25030i$&      &   & $3$ & $0.0000-21.06955i$&$0.25032i$\\
       &   & $4$ &$0.0000-21.32073i$&$0.25032i$&       &    & $4$ & $0.0000-21.32069i$&$0.25032i$&      &   & $4$ & $0.0000-21.31989i$&$0.25034i$\\
       &   & $5$ &$0.0000-21.57104i$&$0.25031i$&       &    & $5$ & $0.0000-21.57100i$&$0.25031i$&      &   & $5$ & $0.0000-21.57022i$&$0.25032i$\\
       &   & $6$ &$0.0000-21.82134i$&$0.25030i$&       &    & $6$ & $0.0000-21.82130i$&$0.25030i$&      &   & $6$ & $0.0000-21.82053i$&$0.25032i$\\
       &   & $7$ &$0.0000-22.07164i$&$0.25030i$&       &    & $7$ & $0.0000-22.07160i$&$0.25030i$&      &   & $7$ & $0.0000-22.07085i$&$0.25032i$\\
       \cline{2-5}
       \cline{7-10}
       \cline{12-15}
       &$2$& $0$ &$0.0000-20.49187i$&$0.25056i$&       & $2$& $0$ & $0.0000-20.49182i$&$0.25057i$&      & $2$  & $0$ & \mbox{N/A}    & \mbox{N/A}\\
       &   & $1$ &$0.0000-20.74263i$&$0.25076i$&       &    & $1$ & $0.0000-20.74257i$&$0.25076i$&      &   & $1$ & \mbox{N/A}    & \mbox{N/A}\\
       &   & $2$ &$0.0000-20.99323i$&$0.25060i$&       &    & $2$ & $0.0000-20.99317i$&$0.25060i$&      &   & $2$ & \mbox{N/A}    & \mbox{N/A}\\
       &   & $3$ &$0.0000-21.24388i$&$0.25065i$&       &    & $3$ & $0.0000-21.24382i$&$0.25069i$&      &   & $3$ & \mbox{N/A}    & \mbox{N/A}\\
       &   & $4$ &$0.0000-21.49452i$&$0.25064i$&       &    & $4$ & $0.0000-21.49446i$&$0.25064i$&      &   & $4$ & \mbox{N/A}    & \mbox{N/A}\\
       &   & $5$ &$0.0000-21.74513i$&$0.25061i$&       &    & $5$ & $0.0000-21.74507i$&$0.25061i$&      &   & $5$ & \mbox{N/A}    & \mbox{N/A}\\
       &   & $6$ &$0.0000-21.99574i$&$0.25062i$&       &    & $6$ & $0.0000-21.99569i$&$0.25062i$&      &   & $6$ & \mbox{N/A}    & \mbox{N/A}\\
       &   & $7$ &$0.0000-22.24634i$&$0.25060i$&       &    & $7$ & $0.0000-22.24629i$&$0.25060i$&      &   & $7$ & \mbox{N/A}    & \mbox{N/A}\\
[1ex]
\hline 
\end{tabular}
\end{table}

\begin{table}
\centering
\caption{\textit{\textcolor{black}{Stable complex spectral roots with $\delta_{\rm R}>0$, denoted $\Omega_{>}$, for $1.30\leq\mu\leq2.20$. Only the $\ell=n=0$ entries are listed. The sign of $\delta_{\rm R}$ places them above the nominal real-frequency threshold but does not establish their Riemann sheet or imply a photon-sphere interpretation in this mass-dominated regime.}}}
\label{tablemu138_14_142_145_150_220}
\setlength\tabcolsep{0.1cm}
\def\arraystretch{1.5}
\begin{tabular}{|c|c|c|}
\hline
\textcolor{black}{$\mu$} & \textcolor{black}{$\Omega_{>}$} & \textcolor{black}{$\delta_{\rm R}$}\\ [0.5ex]
\hline
1.30  & 1.3058-0.3551i & 0.0151 \\
1.40  & 1.4424-0.4156i & 0.1205 \\
1.42  & 1.4424-0.4156i & 0.0641 \\
1.45  & 1.4769-0.4312i & 0.0787 \\
1.50  & 1.5349-0.4576i & 0.1059 \\
2.20  & 2.3906-0.8718i & 0.8750 \\[1ex]
\hline 
\end{tabular}
\end{table}

\textcolor{black}{For clarity, the values $\mu>1$ considered below are large values of a dimensionless coupling, not large particle masses in electronvolts. For the low multipoles used here, this is a mass-dominated regime. Once the condition in Eq.~\eqref{muineq} fails, the usual photon-sphere maximum-minimum structure is absent. We reiterate that we use $\Omega_{<}$ and $\Omega_{>}$ only for roots with $\delta_{\rm R}<0$ and $\delta_{\rm R}>0$, respectively. These symbols do not encode the Riemann sheet or the sign of $\Im {\rm m}\{k\}$, and therefore they do not amount to an evanescent/quasibound versus propagating/outgoing classification. Both groups are retained as numerical tests of the uniformised polynomial problem as the potential becomes step-like.}

\textcolor{black}{Finally, in Tables~\ref{tablemu138_14_142-straight}, \ref{tablemu138_14_142overdamped}, \ref{tablemu145_150_220-straight}, \ref{tablemu145_150_220overdamped}, and \ref{tablemu138_14_142_145_150_220}, we investigate the mass-dominated interval $1.3\leq\mu\leq2.2$. The $\delta_{\rm R}<0$ group dominates the tabulated complex roots, while the $\delta_{\rm R}>0$ group in Table~\ref{tablemu138_14_142_145_150_220} contains only the $\ell=n=0$ entries. In Tables~\ref{tablemu138_14_142-straight} and \ref{tablemu145_150_220-straight}, several $\Omega_{<}$ sequences vary non-monotonically with the tabulation index $n$. For the $\Omega_{>}$, $\ell=n=0$ roots collected in Table~\ref{tablemu138_14_142_145_150_220}, $\Re {\rm e}\{\Omega\}$ increases with $\mu$. These statements concern the displayed frequency sequences only and do not substitute for a sheet-resolved mode labelling. As before, we identify families of purely imaginary stable roots, exhibiting approximately equal spacing between subsequent entries in Tables~\ref{tablemu138_14_142overdamped} and \ref{tablemu145_150_220overdamped}. Unlike previous cases, the retained spectra contain no such roots for $\ell=0$ at $\mu=1.40,1.42$ or for $\ell=2$ at $\mu=2.20$. Their sheet-dependent radial interpretation is not inferred from $\delta_{\rm R}=-\mu^2$.}

Our results reveal {\color{black}{numerically stable families of purely imaginary spectral roots}} in specific regions of parameter space. The numerical claim is that these frequencies pass the three-resolution and increased-precision convergence tests. {\color{black}{Note that while this establishes their stability within the uniformised boundary-value problem, it does not by itself establish them as a distinct physical branch of massive scalar QNMs.}}
 Uniformising $k(\Omega)~=~\sqrt{\Omega^2-\mu^2}$ before discretisation rules out the specific picture in which the massive square-root cut is itself approximated by a finite row of collocation points, and analogous imaginary-axis ladders in the quadratic-pencil calculation of Ref.~\cite{BaticDutykhSukaiti2026KSQC} rule out an explanation specific to the seventh-degree companion form. The present $\Omega$-only archive, however, does not retain the reciprocal pre-image needed to assign each tabulated root to the outgoing or reciprocal sheet.

\textcolor{black}{The weak dependence of the highly damped branch on $\mu$ is also not anomalous. On the two sheets and for $|\Omega|\gg\mu$,}
\begin{equation}\label{eq:large-damping-dispersion}
\textcolor{black}{k_{\pm}(\Omega)=\pm\sqrt{\Omega^2-\mu^2}=\pm\left[\Omega-\frac{\mu^2}{2\Omega}+\mathcal{O}\!\left(\frac{\mu^4}{\Omega^3}\right)\right].}
\end{equation}
\textcolor{black}{For $\Omega=-i\Gamma$ with $\Gamma\simeq 20$, the relative mass correction is of order $\mu^2/\Gamma^2$ on either sheet, so weak mass dependence is expected. Likewise, the elementary spacing $\Delta(-\Im {\rm m}\{\Omega\})\simeq 0.25$ coincides with the natural Schwarzschild scale $M\kappa=1/4$. We regard this agreement as structural evidence for an overdamped ladder, not as an analytic proof of its exact asymptotic formula.}

\textcolor{black}{The additional diagnostics proposed for ordinary weakly damped ringdown modes are not equally probative here. A plot of the unfactored QNM radial function is generally unbounded at an asymptotic endpoint for a damped outgoing resonance, whereas a plot of the factored function $\Phi$ merely displays the regularity already enforced by the ansatz and does not test the Riemann-sheet choice independently. A time-domain non-detection would also not falsify these roots. For the first tabulated damping rate, $-\Im {\rm m}\{\Omega\}\simeq 19.37$, the e-folding time is only $\tau/M\simeq 0.052$, and the amplitude is suppressed by $e^{-19.37}\simeq 3.9\times10^{-9}$ already at $t=M$. Such a contribution is buried in the prompt response long before a conventional ringdown fit is performed, and its visibility additionally depends on the excitation residue. A direct construction of the retarded Green function, including pole residues and the discontinuity across the massive-field branch cut, followed by the inverse-frequency transform, would be required to determine the pole-versus-continuum decomposition and the late-time signal. This is a separate problem from establishing numerically stable roots of the boundary-value problem.}

\textcolor{black}{As illustrated in Section~\ref{sec:context}, introducing a mass term changes the far-region structure of the effective potential and the analytic boundary condition. The dependence of the roots on $\mu$ expresses sensitivity to the mass of the perturbing field and to the far-region effective potential. It is not an additional property or \emph{hair} of the Schwarzschild black hole. The numerical tables show how the stable roots vary with $\ell$, $n$, and $\mu$. For small and intermediate couplings, the usual weakly damped branches evolve continuously from the massless benchmarks. At larger couplings, roots with $\delta_{\rm R}<0$ dominate the displayed data, but we no longer call this an evanescent sector because the spatial behaviour requires $k(\Lambda)$ and its sheet.}

\textcolor{black}{It is useful to revisit earlier claims concerning the highly damped limit of a massive scalar field. Hod observed that $\Re {\rm e}\{\Omega\}\sim\ln 3/(8\pi M)$ as $n\to\infty$ for Schwarzschild perturbations \cite{refNollert1993} and interpreted this value in connection with black hole entropy and area quantisation \cite{Hod1998PRL}. Motl subsequently derived the $\ln{3}$ result for massless scalar and gravitational perturbations by applying monodromy methods to the radial equation in the complex plane \cite{refMotl}. Ref.~\cite{Konoplya2005PLB} argued that the same leading behaviour persists for a massive scalar field by expanding the coefficients of the Leaver-type three-term recurrence in the large-$|\omega|$ limit. At leading order, the recurrence reduces to the massless form up to a shift of the effective multipole number. Because the massless high-overtone asymptotics are independent of $\ell$, this reduction was taken to imply the same $\ln 3/(8\pi M)$ limit for every finite $\mu$. The argument therefore imports the massless analytic structure through an expansion that is formally valid in the large-frequency regime.}

\textcolor{black}{The complete stable root set produced by the uniformised problem does not show the real part approaching $\ln 3/(8\pi M)$ over the resolved range when $\mu\neq 0$. This observation warrants caution, but it does not by itself disprove an asymptotic statement restricted to the outgoing sheet. What can be stated is that the full two-sheeted spectrum is not described by a naive import of the massless pattern. The non-vanishing asymptotic value of the potential changes the global analytic structure of the radial equation and its Stokes geometry, so arguments that rely on a vanishing potential at infinity, as in the monodromy construction of Ref.~\cite{refMotl}, do not straightforwardly generalise to every branch of the massive problem.}

\textcolor{black}{This observation has a more limited but still important implication: the stable highly damped roots of the uniformised two-sheeted problem remain sensitive to the far-region structure and are not determined solely by near-horizon physics. It also emphasises the restricted domain of analytic continuations that reduce the massive recurrence relation to its massless form. Because the present archive does not retain the sheet of every root, the absence of the $\ln 3/(8\pi M)$ trend should at this stage be read as a statement about the complete stable root set produced by the uniformised problem, not as a final theorem about the outgoing QNM sheet alone.}

\textcolor{black}{The branch-resolved recurrence calculation in Table~\ref{table:direct-massive-benchmark} supplies a quantitative check for selected near-threshold roots. For $\mu=0.7$ with $\ell=0$, and for $\mu=0.8$ with $\ell=0,1$, the independently computed spatially decaying continued-fraction solutions agree with the archived spectral frequencies at the $10^{-17}$ level. This establishes that those complex frequencies occur in the decaying recurrence problem and provides a direct check of the quasiresonant/quasibound side of the calculation. The comparison also sharpens the domains of the standard methods. WKB remains appropriate for low-lying barrier-controlled modes but loses its organising turning-point structure as the barrier is suppressed. Continued fractions do not generically fail for a massive field. Indeed,  Tables~\ref{table:literature-massive-benchmark} and \ref{table:direct-massive-benchmark} show that a branch-resolved Leaver-Nollert calculation can be extremely accurate. Its numerical conditioning becomes delicate near $k=0$, where inverse powers of $k$ enter the asymptotic exponent and recurrence, and branch tracking is essential. For the large-$\mu$ stress tests and the highly damped imaginary-axis branches, for which we have not found a pointwise external benchmark, we therefore report convergence of the uniformised boundary-value roots without claiming independent confirmation of their physical sheet or excitation.}

\textcolor{black}{The extended spectral formulation is consequently best viewed as complementary to recurrence and WKB methods. It does not require WKB barrier matching, treats the two-sheeted dispersion relation before discretisation, and provides a global root catalogue. Its reliability in the overlap domain is now demonstrated by massive-field agreement with an independent recurrence calculation, while its extrapolation beyond that domain is supported by three-resolution and precision stability. Numerical convergence of $\Omega$, Riemann-sheet assignment, and dynamical relevance nevertheless remain separate tests.}


\section{Conclusions and outlook \label{sec:conc}}

\textcolor{black}{In this work, we revisited the massive scalar spectral problem on a Schwarzschild background. Given that WKB techniques are most effective for barrier-controlled modes, and whilst continued fractions remain powerful provided that the massive asymptotic branch is fixed and the recurrence is well conditioned, a more robust approach is desired. As such, we extended the Chebyshev spectral framework of Refs.~\cite{Batic2024CQG, Batic2024PRD, Batic2025EPJC, Batic2025CQG, Batic2025PRSA, Batic2026EPJC} by uniformising the massive dispersion relation and incorporating its two-sheeted asymptotic factors into the radial ansatz. From there we validated the matrix construction against established massless results in Table~\ref{tablemu0} and, more importantly, against massive-field Hill-determinant and Leaver-Nollert benchmarks in Tables~\ref{table:literature-massive-benchmark} and \ref{table:direct-massive-benchmark}. The direct comparison covers fundamental modes, selected overtones, and low-damping decaying solutions. The largest displayed difference is $1.5\times10^{-7}$, with all other rows agreeing to approximately $10^{-15}$ or better. The method then gives stable roots for $\ell<n$ and for both signs of the nominal threshold proxy $\delta_{\rm R}$. For complex frequencies, this proxy is not a physical propagating/evanescent classification; the exact radial character requires the complex $k(\Lambda)$ and the Riemann sheet.}

\textcolor{black}{We emphasise that these minimally coupled scalar frequencies are not intended as direct templates for the tensorial Kerr ringdown targeted by gravitational-wave spectroscopy. The present calculation is a controlled first step aimed at validating the treatment of massive-field asymptotics and branch structure. The principal observational extensions are the gravitational perturbation problem for Kerr backgrounds and, within concrete particle physics or beyond-GR models, coupled scalar, vector, or tensor sectors together with an explicit excitation and detection mechanism.}

\textcolor{black}{The dimensionless range explored here should likewise not be read as a generic statement of particle physics viability. Its conversion to a physical mass depends inversely on the black hole mass, as shown in Table~\ref{table:physical_mass_conversion}, and present superradiance constraints are conditional on the field model and the astrophysical system. The cases $\mu\in\{3,10,100\}$ in Appendix~A are retained only as numerical stress tests. At the low multipoles studied there, the potential is mass dominated. $\Omega_{>}$ and $\Omega_{<}$ refer only to the sign of $\delta_{\rm R}$ and carry no unverified sheet or photon-sphere interpretation.}

A particularly interesting outcome is the {\color{black}{emergence of numerically stable families of purely imaginary spectral roots}} with approximately constant spacing in $\Im {\rm m}\{\Omega\}$. We emphasise that we do not identify these roots as physical QNMs: numerical convergence establishes their stability within the uniformised spectral problem, but the present $\Omega$-only archive does not determine their sheet assignment. The individual-resolution data in Table~\ref{table:imaginary-convergence} and the full convergence reports rule out a simple resolution-drift explanation, while the occurrence of analogous ladders in the quadratic-pencil calculation of Ref.~\cite{BaticDutykhSukaiti2026KSQC} provides additional evidence against an artefact specific to the seventh-degree companion linearisation. The near-threshold $\delta_{\rm R}<0$ cluster in Fig.~\ref{fig:QNMplot} is distinct from the highly damped purely imaginary ladder. The same figure also shows local spectral clustering of the type considered in defective-cluster theory \cite{TisseurMeerbergen,Beyn2014}: as the truncation order $N$ increases, a group of numerically stable roots accumulates near $\Omega=\mu+0i$. A single discrete spectrum cannot distinguish rigorously between a genuinely defective continuous spectral point and several tightly packed semisimple roots. The observed behaviour is nevertheless compatible with the splitting expected near an isolated defective or nearly defective component. This interpretation concerns local clustering only and, like the convergence test, does not replace the missing per-root sheet information.

\textcolor{black}{The stable high-damping spectrum obtained from the uniformised problem is sensitive to the global structure of the effective potential. Within the resolved two-sheeted root set, we find no approach to $\ln 3/(8\pi M)$. This is evidence that the nonzero asymptote matters, but it should not be promoted to a universal statement about the outgoing QNM sheet until the retained $\Lambda$ values have been used to separate the two sheets.}

\textcolor{black}{The scope of these conclusions is spectral. We have not constructed the retarded Green function, evaluated its branch-cut integral, or evolved initial data in time, and we therefore draw no conclusion about late-time tail exponents, amplitudes, or crossover times. Once a root has been assigned to the appropriate sheet and boundary condition, a small negative $\Im {\rm m}\{\Omega\}$ does define a long decay time for that discrete mode. It does not by itself determine the excitation amplitude or whether the mode dominates over the continuum contribution. Likewise, the dependence on $\mu$ reflects sensitivity to the perturbing field mass and the far-zone potential, not black hole hair.}

\textcolor{black}{The extended method can be applied to charged or rotating black holes, with or without a cosmological constant, where superradiance, mass thresholds, and additional sheets may enrich the spectrum. In such applications, the numerical output should preserve the uniformising variable and the complex wavenumber so that outgoing-resonant and decaying quasibound sectors are classified directly rather than through a real-frequency proxy. It would also be valuable to seek an analytic interpretation of the {\color{black}{numerically stable purely imaginary spectral families}} through a suitable generalisation of monodromy, continued-fraction, Wronskian, or resolvent techniques.}

\section*{Code availability}

\par The repository, Ref.~\cite{Batic2026_MassiveScalar}, contains the raw $\Omega$ spectra at $N\in\{180,190,200\}$, the parameter-by-parameter convergence reports, and the matrix-assembly and polynomial-eigenvalue routines used for the tables. The original archived files contain only $\Re {\rm e} \{\Omega\}$ and $\Im {\rm m}\{\Omega\}$. The revised solver supplied with this revision additionally exports $\Lambda$, $k(\Lambda)$, and explicit sheet/asymptotic indicators so that future tables can be filtered without ambiguity. The accompanying reproducibility files also contain the independent Leaver-Nollert benchmark script, its branch prescription, continued-fraction truncation sequence, and the numerical output used in Tables~\ref{table:literature-massive-benchmark} and \ref{table:direct-massive-benchmark}.

\newpage
\appendix

\section{\texorpdfstring{\textcolor{black}{Large-$\mu$ numerical stress tests}}{Large-mu numerical stress tests} \label{app:A}}

\par \textcolor{black}{The values $\mu\in\{3,10,100\}$ in this appendix are included solely to test the numerical formulation at large dimensionless coupling; they are not presented as generically viable or observationally preferred particle masses. Eq.~\eqref{eq:physical_mu} gives $\lambda_C/r_s=1/6$, $1/20$, and $1/200$, respectively. For $\ell\leq2$, the effective potential is mass dominated and no longer resembles the usual photon-sphere barrier. The symbols $\Omega_{>}$ and $\Omega_{<}$ distinguish only the signs of $\delta_{\rm R}$. The current $\Omega$-only archive does not support a per-root outgoing, reciprocal, growing, decaying, or quasibound label. The value of these cases is methodological: they test the convergence and dynamic range of the uniformised polynomial problem.}

\par \textcolor{black}{In Tables~\ref{tablelargemu} and \ref{tablelargemuoverdamped}, we explore $\mu=3,10,100$ at low multipoles. For $\mu=3$, the $\delta_{\rm R}>0$ group contains two entries at $\ell=0$, one at $\ell=1$, and none at $\ell=2$; for $\mu=10$, it contains several entries for all three multipoles; and for $\mu=100$, only a small $\delta_{\rm R}<0$ subset is resolved. The $\Omega_{<}$ sequences often vary non-monotonically with $n$. These counts describe the neutral threshold grouping only and should not be read as counts of propagating QNMs versus evanescent modes.}

\par \textcolor{black}{In Table~\ref{tablelargemuoverdamped}, we again find purely imaginary stable roots, with $|\Im {\rm m}\{\Omega\}|$ increasing along the retained sequences. For $\mu=3$, the spacing is approximately $0.25004i$ from the first displayed entries; for higher multipoles and for $\mu=10$, the spacing approaches an approximately constant value as the tabulation index increases. No corresponding stable imaginary sequence is resolved at $\mu=100$ within the present numerical settings. These statements concern convergence and spacing only, not sheet assignment.}

\par \textcolor{black}{As discussed in Section~\ref{sec:results}, Ref.~\cite{Konoplya2005PLB} argues that $\Re {\rm e}\{\Omega\}\to\ln 3/(8\pi)$ as $n\to\infty$, irrespective of whether the dimensionless coupling is small or large; that study considered $\mu=0.3$ and $\mu=3$ for $\ell=0$.} 
\textcolor{black}{Our spectral simulations do not show this behaviour in the complete stable root set. For $\mu\gtrsim3$, the real part of several retained complex-root sequences varies non-monotonically with the tabulation index: it initially decreases, reaches a minimum, and then increases again, with no indication of convergence toward $\ln 3/(8\pi)$ over the reliably resolved range. A statement restricted to the outgoing QNM sheet requires the sheet-resolved output described above.}

\begin{table}[ht]
\centering
\caption{\textit{\textcolor{black}{Stable complex spectral roots in the large-$\mu$ numerical stress tests. The columns $\Omega_{>}$ and $\Omega_{<}$ contain roots with $\delta_{\rm R}>0$ and $\delta_{\rm R}<0$, respectively. These symbols are neutral table groupings and do not establish outgoing/reciprocal-sheet membership or resonant/quasibound asymptotics. \enquote{N/A} indicates data not available.}}}
\label{tablelargemu}
\vspace*{1em}
\def\arraystretch{1.5}
\begin{tabular}{@{}|c|c|c|c|c||c|c|c|c|c||c|c|c|c|c|c|@{}}
\hline
\textcolor{black}{$\mu$} & \textcolor{black}{$\ell$} & \textcolor{black}{$n$} & \textcolor{black}{$\Omega_{>}$} & \textcolor{black}{$\Omega_{<}$} & \textcolor{black}{$\mu$} & \textcolor{black}{$\ell$} & \textcolor{black}{$n$} & \textcolor{black}{$\Omega_{>}$} & \textcolor{black}{$\Omega_{<}$} & \textcolor{black}{$\mu$} & \textcolor{black}{$\ell$} & \textcolor{black}{$n$} & \textcolor{black}{$\Omega_{>}$} & \textcolor{black}{$\Omega_{<}$}\\ [0.5ex]
\hline
$3$&$0$&$0$&$3.0842-0.6922i$&$2.9305-0.1137i$&$10$&$0$&$0$&$10.0683-1.6430i$&$9.8058-0.9043i$&$10^2$&$0$&$0$&\mbox{N/A}&$97.9653-8.1301i$\\
   &   &$1$&$3.4567-1.4295i$&$2.9269-0.1426i$&    &   &$1$&$10.1768-1.8563i$&$9.8249-0.9880i$&      &   &$1$&\mbox{N/A}&\mbox{N/A}\\
   &   &$2$&\mbox{N/A}      &$2.9254-0.1820i$&    &   &$2$&$10.3277-2.1214i$&$9.8504-1.0832i$&      &   &$2$&\mbox{N/A}&\mbox{N/A}\\
   &   &$3$&\mbox{N/A}      &$2.9287-0.2373i$&    &   &$3$&$10.5448-2.4628i$&$9.8843-1.1924i$&      &   &$3$&\mbox{N/A}&\mbox{N/A}\\
   &   &$4$&\mbox{N/A}      &$2.9425-0.3187i$&    &   &$4$&$10.8737-2.9265i$&$9.9291-1.3189i$&      &   &$4$&\mbox{N/A}&\mbox{N/A}\\
   &   &$5$&\mbox{N/A}      &$2.9803-0.4489i$&    &   &$5$&$11.4156-3.6145i$&$9.9886-1.4670i$&      &   &$5$&\mbox{N/A}&\mbox{N/A}\\
   \cline{2-5}
   \cline{7-10}
   \cline{12-15}
   &$1$&$0$&$3.1422-1.1956i$&$2.9220-0.1341i$&    &$1$&$0$&$10.0419-1.6272i$&$9.7974-0.8961i$&      &$1$&$0$&\mbox{N/A}&\mbox{N/A}\\
   &   &$1$&\mbox{N/A}      &$2.9177-0.1710i$&    &   &$1$&$10.1429-1.8380i$&$9.8151-0.9790i$&      &   &$1$&\mbox{N/A}&\mbox{N/A}\\
   &   &$2$&\mbox{N/A}      &$2.9163-0.2227i$&    &   &$2$&$10.2829-2.0998i$&$9.8387-1.0733i$&      &   &$2$&\mbox{N/A}&\mbox{N/A}\\
   &   &$3$&\mbox{N/A}      &$2.9213-0.2985i$&    &   &$3$&$10.4830-2.4364i$&$9.8702-1.1814i$&      &   &$3$&\mbox{N/A}&\mbox{N/A}\\
   &   &$4$&\mbox{N/A}      &$2.9408-0.4184i$&    &   &$4$&$10.7828-2.8923i$&$9.9119-1.3066i$&      &   &$4$&\mbox{N/A}&\mbox{N/A}\\
   &   &$5$&\mbox{N/A}      &$2.9961-0.6359i$&    &   &$5$&$11.2669-3.5652i$&$9.9675-1.4531i$&      &   &$5$&\mbox{N/A}&\mbox{N/A}\\
      \cline{2-5}
   \cline{7-10}
   \cline{12-15}
   &$2$&$0$&\mbox{N/A}      &$2.9069-0.1546i$&    &$2$&$0$&$10.0847-1.8111i$&$9.7973-0.9651i$&      &$2$&$0$&\mbox{N/A}&$97.9490-8.0581i$\\
   &   &$1$&\mbox{N/A}      &$2.8992-0.2010i$&    &   &$1$&$10.2069-2.0683i$&$9.8177-1.0582i$&      &   &$1$&\mbox{N/A}&$97.9608-8.1252i$\\
   &   &$2$&\mbox{N/A}      &$2.8931-0.2684i$&    &   &$2$&$10.3796-2.3983i$&$9.8453-1.1646i$&      &   &$2$&\mbox{N/A}&$97.9730-8.1931i$\\
   &   &$3$&\mbox{N/A}      &$2.8910-0.3730i$&    &   &$3$&$10.6335-2.8436i$&$9.8818-1.2879i$&      &   &$3$&\mbox{N/A}&$97.9856-8.2617i$\\
   &   &$4$&\mbox{N/A}      &$2.8961-0.5539i$&    &   &$4$&$11.0298-3.4948i$&$9.9306-1.4322i$&      &   &$4$&\mbox{N/A}&\mbox{N/A}\\
   &   &$5$&\mbox{N/A}      &$2.8910-0.9412i$&    &   &$5$&$11.7102-4.6002i$&$9.9961-1.6036i$&      &   &$5$&\mbox{N/A}&\mbox{N/A}\\
[1ex]
\hline
\end{tabular}
\end{table}

\begin{table}[ht]
\centering
\caption{\textit{\textcolor{black}{Purely imaginary stable spectral roots for $\mu\in\{3,10,100\}$ and $\ell\in\{0,1,2\}$. The results use $200$ polynomials and $200$-digit precision, with $\Delta\Omega=\Omega_n-\Omega_{n+1}$. All entries have $\delta_{\rm R}=-\mu^2$, but this does not determine the Riemann sheet or the sign of $\Im {\rm m}\{k\}$. Entries labelled \enquote{N/A} indicate unavailable data.}}}
\label{tablelargemuoverdamped}
\vspace*{1em}
\def\arraystretch{1.5}
\begin{tabular}{@{}|c|c|c|c|c||c|c|c|c|c||c|c|c|c|c|@{}}
\hline
$\mu$  &$\ell$ & $n$ &$\Omega$         & $\Delta\Omega$  & $\mu$   & $\ell$ & $n$ & $\Omega$       & $\Delta\Omega$ & $\mu$   & $\ell$ & $n$ & $\Omega$ & $\Delta\Omega$ \\ [0.5ex]
\hline
$3$    &$0$& $0$ &$0.0000-31.36718i$&$0.25004i$&$10$&$0$&$0$&$0.0000-22.54378i$&$0.25222i$&$10^2$&$0$&$0$&\mbox{N/A}&\mbox{N/A}\\
       &   & $1$ &$0.0000-31.61723i$&$0.25004i$&    &   &$1$&$0.0000-22.79587i$&$0.25209i$&      &   &$1$&\mbox{N/A}&\mbox{N/A}\\
       &   & $2$ &$0.0000-31.86727i$&$0.25004i$&    &   &$2$&$0.0000-23.04794i$&$0.25208i$&      &   &$2$&\mbox{N/A}&\mbox{N/A}\\
       &   & $3$ &$0.0000-32.11731i$&$0.25004i$&    &   &$3$&$0.0000-23.29990i$&$0.25196i$&      &   &$3$&\mbox{N/A}&\mbox{N/A}\\
       &   & $4$ &$0.0000-32.36735i$&$0.25004i$&    &   &$4$&$0.0000-23.55177i$&$0.25187i$&      &   &$4$&\mbox{N/A}&\mbox{N/A}\\
       &   & $5$ &$0.0000-32.61739i$&$0.25004i$&    &   &$5$&$0.0000-23.80358i$&$0.25180i$&      &   &$5$&\mbox{N/A}&\mbox{N/A}\\
       &   & $6$ &$0.0000-32.86743i$&$0.25004i$&    &   &$6$&$0.0000-24.05530i$&$0.25173i$&      &   &$6$&\mbox{N/A}&\mbox{N/A}\\
       &   & $7$ &$0.0000-33.11747i$&$0.25004i$&    &   &$7$&$0.0000-24.30696i$&$0.25166i$&      &   &$7$&\mbox{N/A}&\mbox{N/A}\\
          \cline{2-5}
   \cline{7-10}
   \cline{12-15}
       &$1$& $0$ &$0.0000-20.06658i$&$0.25058i$&    &$1$&$0$&$0.0000-22.99102i$&$0.25289i$&      &   &$0$&\mbox{N/A}&\mbox{N/A}\\
       &   & $1$ &$0.0000-20.31684i$&$0.25025i$&    &   &$1$&$0.0000-23.24376i$&$0.25273i$&      &   &$1$&\mbox{N/A}&\mbox{N/A}\\
       &   & $2$ &$0.0000-20.56729i$&$0.25046i$&    &   &$2$&$0.0000-23.49632i$&$0.25257i$&      &   &$2$&\mbox{N/A}&\mbox{N/A}\\
       &   & $3$ &$0.0000-20.81767i$&$0.25037i$&    &   &$3$&$0.0000-23.74882i$&$0.25250i$&      &   &$3$&\mbox{N/A}&\mbox{N/A}\\
       &   & $4$ &$0.0000-21.06803i$&$0.25036i$&    &   &$4$&$0.0000-24.00121i$&$0.25239i$&      &   &$4$&\mbox{N/A}&\mbox{N/A}\\
       &   & $5$ &$0.0000-21.31841i$&$0.25038i$&    &   &$5$&$0.0000-24.25350i$&$0.25229i$&      &   &$5$&\mbox{N/A}&\mbox{N/A}\\
       &   & $6$ &$0.0000-21.56876i$&$0.25035i$&    &   &$6$&$0.0000-24.50570i$&$0.25220i$&      &   &$6$&\mbox{N/A}&\mbox{N/A}\\
       &   & $7$ &$0.0000-21.81911i$&$0.25035i$&    &   &$7$&$0.0000-24.75782i$&$0.25212i$&      &   &$7$&\mbox{N/A}&\mbox{N/A}\\
          \cline{2-5}
   \cline{7-10}
   \cline{12-15}
       &$2$& $0$ &$0.0000-20.73928i$&$0.25082i$&    &$2$&$0$&$0.0000-22.90000i$&$0.25333i$&      &   &$0$&\mbox{N/A}&\mbox{N/A}\\
       &   & $1$ &$0.0000-20.98992i$&$0.25064i$&    &   &$1$&$0.0000-23.15334i$&$0.25334i$&      &   &$1$&\mbox{N/A}&\mbox{N/A}\\
       &   & $2$ &$0.0000-21.24068i$&$0.25076i$&    &   &$2$&$0.0000-23.40641i$&$0.25307i$&      &   &$2$&\mbox{N/A}&\mbox{N/A}\\
       &   & $3$ &$0.0000-21.49137i$&$0.25069i$&    &   &$3$&$0.0000-23.65942i$&$0.25301i$&      &   &$3$&\mbox{N/A}&\mbox{N/A}\\
       &   & $4$ &$0.0000-21.74205i$&$0.25068i$&    &   &$4$&$0.0000-23.91236i$&$0.25294i$&      &   &$4$&\mbox{N/A}&\mbox{N/A}\\
       &   & $5$ &$0.0000-21.99274i$&$0.25069i$&    &   &$5$&$0.0000-24.16518i$&$0.25282i$&      &   &$5$&\mbox{N/A}&\mbox{N/A}\\
       &   & $6$ &$0.0000-22.24340i$&$0.25066i$&    &   &$6$&$0.0000-24.41791i$&$0.25273i$&      &   &$6$&\mbox{N/A}&\mbox{N/A}\\
       &   & $7$ &$0.0000-22.49405i$&$0.25065i$&    &   &$7$&$0.0000-24.67056i$&$0.25265i$&      &   &$7$&\mbox{N/A}&\mbox{N/A}\\
[1ex]
\hline 
\end{tabular}
\end{table}

\section*{Acknowledgements}
AC acknowledges support from the Initiative Physique des Infinis (IPI), a research-training programme of IdEx SUPER at Sorbonne Universit\'e. ASC was partly supported by the National Research Foundation of South Africa and thanks New York University Abu Dhabi for its hospitality during the completion of this work.

\bibliography{QNMS}

@misc{Batic2026_MassiveScalar,
  author       = {Davide Batic and Anna Chrysostomou and Alan S. Cornell and Denys Dutykh},
  title        = {Massive Scalar Field in a Schwarzschild Background},
  year         = {2026},
  howpublished = {\url{https://github.com/dutykh/massive-scalar}},
  note         = {GitHub repository}
}

@article{TisseurMeerbergen,
author = {Tisseur, Fran\c{c}oise and Meerbergen, Karl},
title = {The Quadratic Eigenvalue Problem},
journal = {SIAM Review},
volume = {43},
number = {2},
pages = {235-286},
year = {2001},
doi = {10.1137/S0036144500381988},
URL = {https://doi.org/10.1137/S0036144500381988},
eprint = {https://doi.org/10.1137/S0036144500381988},
}

@article{Beyn2014,
author = {Beyn, Wolf-Jürgen and Latushkin, Yuri and Rottmann-Matthes, Jens},
title = {Finding Eigenvalues of Holomorphic Fredholm Operator Pencils Using Boundary Value Problems and Contour Integrals},
journal = {Integral Equations and Operator Theory},
volume = {78},
number = {2},
pages = {155–211},
year = {2014},
doi = {10.1007/s00020-013-2117-6},
URL = {https://doi.org/10.1007/s00020-013-2117-6},
}

@article{Konoplya:2003ii,
    author = "Konoplya, R. A.",
    title = "{Quasinormal behavior of the d-dimensional Schwarzschild black hole and higher order WKB approach}",
    eprint = "gr-qc/0303052",
    archivePrefix = "arXiv",
    doi = "10.1103/PhysRevD.68.024018",
    journal = "Phys. Rev. D",
    volume = "68",
    pages = "024018",
    year = "2003"
}

@article{refNollert1993,
    author = "Nollert, Hans-Peter",
    title = "{Quasinormal modes of Schwarzschild black holes: The determination of quasinormal frequencies with very large imaginary parts}",
    doi = "10.1103/PhysRevD.47.5253",
    journal = "Phys. Rev. D",
    volume = "47",
    pages = "5253--5258",
    year = "1993"
}

@article{refMotl,
    author = "Motl, Lubo{\v{s}}",
    title = "{An Analytical computation of asymptotic Schwarzschild quasinormal frequencies}",
    eprint = "gr-qc/0212096",
    archivePrefix = "arXiv",
    reportNumber = "HEP-UK-0016, HUTP-02-A066",
    doi = "10.4310/ATMP.2002.v6.n6.a3",
    journal = "Adv. Theor. Math. Phys.",
    volume = "6",
    pages = "1135--1162",
    year = "2003",
}

@article{Fortuna:2020obg,
    author = "Fortuna, Sean and Vega, Ian",
    title = "{Bernstein spectral method for quasinormal modes and other eigenvalue problems}",
    eprint = "2003.06232",
    archivePrefix = "arXiv",
    primaryClass = "gr-qc",
    doi = "10.1140/epjc/s10052-023-12350-9",
    journal = "Eur. Phys. J. C",
    volume = "83",
    number = "12",
    pages = "1170",
    year = "2023"
}

@article{Vishveshwara1970_scattering,
  title = {Scattering of Gravitational Radiation by a {{Schwarzschild}} Black-Hole},
  author = {Vishveshwara, C V},
  year = {1970},
  journal = {Nature},
  volume = {227},
  pages = {936--938},
  doi = {10.1038/227936a0}
}

@article{Vishveshwara1970_stability,
  title = {Stability of the {{Schwarzschild}} Metric},
  author = {Vishveshwara, C. V.},
  year = {1970},
  journal = {Phys. Rev. D},
  volume = {1},
  pages = {2870--2879},
  doi = {10.1103/PhysRevD.1.2870}
}

@article{Press1971,
  title = {Long Wave Trains of Gravitational Waves from a Vibrating Black Hole},
  author = {Press, William H.},
  year = {1971},
  journal = {Astrophys. J. Lett.},
  volume = {170},
  number = {OAP-262},
  pages = {L105--L108},
  doi = {10.1086/180849}
}

@article{Dolan2007_MassiveScalarKerr,
  title = {Instability of the Massive {{Klein-Gordon}} Field on the {{Kerr}} Spacetime},
  author = {Dolan, Sam R.},
  year = {2007},
  journal = {Phys. Rev. D},
  volume = {76},
  eprint = {0705.2880},
  primaryclass = {gr-qc},
  pages = {084001},
  doi = {10.1103/PhysRevD.76.084001},
  archiveprefix = {arxiv}
}

@article{Dolan2020_MassKerr,
  title = {Quasinormal Modes of Massive Vector Fields on the {{Kerr}} Spacetime},
  author = {Percival, Jake and Dolan, Sam R.},
  year = {2020},
  journal = {Phys. Rev. D},
  volume = {102},
  number = {10},
  eprint = {2008.10621},
  primaryclass = {gr-qc},
  pages = {104055},
  doi = {10.1103/PhysRevD.102.104055},
  archiveprefix = {arxiv}
}

@article{Konoplya2024_TwoRegimes,
  title = {Two Regimes of Asymptotic Fall-off of a Massive Scalar Field in the {{Schwarzschild}}--de {{Sitter}} Spacetime},
  author = {Konoplya, R. A.},
  year = 2024,
  journal = {Phys. Rev. D},
  volume = {109},
  number = {10},
  eprint = {2401.17106},
  primaryclass = {gr-qc},
  pages = {104018},
  doi = {10.1103/PhysRevD.109.104018},
  archiveprefix = {arXiv}
}

@article{SimoneWill1991_MassiveScalar,
  title = {Massive Scalar Quasinormal Modes of {{Schwarzschild}} and {{Kerr}} Black Holes},
  author = {Simone, Liliana E. and Will, Clifford M.},
  year = 1992,
  journal = {Class. Quant. Grav.},
  volume = {9},
  number = {WUGRAV-91-4},
  pages = {963--978},
  doi = {10.1088/0264-9381/9/4/012}
}

@article{Arvanitaki2010PRD, 
  title={String axiverse},
  author={A. Arvanitaki and S. Dimopoulos and S. Dubovsky and N. Kaloper and J. March-Russell},
  journal={Phys. Rev. D},
  volume={81},
  pages={123530},
  year={2010}
}

@article{Batic2024CQG, 
  title={A Unified Spectral Approach for Quasinormal Modes of Morris-Thorne Wormholes},
  author={D. Batic and D. Dutykh},
  journal={Class. Quantum Grav.},
  volume={41},
  pages={215003},
  year={2024}
}

@article{Batic2024EPJC, 
  title={Quasinormal Modes in Noncommutative Schwarzschild Black Holes: A
  Spectral Analysis},
  author={D. Batic and D. Dutykh},
  journal={Eur. Phys. J. C},
  volume={84},
  pages={622},
  year={2024}
}

@article{Batic2024PRD, 
  title={A Unified Spectral Approach for Quasinormal Modes of Lee-Wick Black Holes},
  author={D. Batic and D. Dutykh and B. Loureiro Giacchini},
  journal={Phys. Rev. D},
  volume={110},
  pages={084032},
  year={2024}
}

@article{Batic2025EPJC, 
  title={Instability analysis of massive static phantom wormholes via the spectral method},
  author={D. Batic and D. Dutykh},
  journal={Eur. Phys. J. C},
  volume={85},
  pages={144},
  year={2025}
}

@article{Batic2025CQG, 
  title={A spectral approach for quasinormal frequencies of noncommutative geometry-inspired wormholes},
  author={D. Batic and D. Dutykh and J. J. Beek},
  journal={Class. Quantum Grav.},
  volume={42},
  pages={085003},
  year={2025}
}

@article{Batic2026EPJC, 
  title={Spectral Analysis of Quasinormal Modes of Planck Stars},
  author={D. Batic and D. Dutykh and F. Scardigli},
  journal={Eur. Phys. J. C},
  volume={86},
  pages={165},
  year={2026}
}

@article{Batic2025PRSA, 
  title={Quasi-normal modes of non-commutative geometry-inspired dirty black holes},
  author={D. Batic and D. Dutykh and Z. Ahmed Babou},
  journal={Proc. R. Soc. Lond. A},
  volume={481},
  pages={20250021},
  year={2025}
}

@article{Berti2009CQG, 
  title={Quasinormal modes of black holes and black branes},
  author={E. Berti and V. Cardoso and A. O Starinets},
  journal={Class. Quantum Grav.},
  volume={26},
  pages={163001},
  year={2009}
}

@article{Hod1998PRL, 
  title={Bohr’s correspondence principle and the area spectrum of quantum black holes},
  author={S. Hod},
  journal={Phys. Rev. Lett.},
  volume={81},
  pages={4293},
  year={1998}
}

@article{Joukowsky1910ZFM, 
  title={Über die Konturen der Tragflächen der Drachenflieger},
  author={N. E. Joukowsky},
  journal={Zeitschrift für Flugtechnik und Motorluftschiffahrt},
  volume={1},
  pages={281},
  year={1910}
}

@article{Konoplya2005PLB, 
  title={Decay of massive scalar field in a Schwarzschild background},
  author={R. A. Konoplya and A. V. Zhidenko},
  journal={Phys. Lett. B},
  volume={609},
  pages={377},
  year={2005}
}

@article{Leaver1985PRSLA, 
  title={An Analytic Representation for the Quasi-Normal Modes of Kerr Black Holes},
  author={E. W. Leaver},
  journal={Proc. R. Soc. Lond. A},
  volume={402},
  number={},
  pages={285},
  year={1985}
}

@article{Mamani2022EPJC, 
  title={Revisiting the quasinormal modes of the Schwarzschild black hole: Numerical analysis},
  author={L. A. H. Mamani and A. D. D. Masa and L. T. Sanches and V. T. Zanchin},
  journal={Eur. Phys. J. C},
  volume={82},
  pages={897},
  year={2022},
  publisher={Springer}
}

@article{Ohashi2004CQG,
  title={Massive quasi-normal mode},
  author={A. Ohashi and M. Sakagami},
  journal={Class. Quant. Grav.},
  volume={21},
  pages={3973},
  year={2004}
}

@article{Olver1994MAA,
  author  = {Olde Daalhuis, A. B. and Olver, F. W. J.},
  title   = {On the calculation of {S}tokes multipliers for linear differential equations of the second order},
  journal = {Methods and Applications of Analysis},
  year    = {1994},
  volume  = {1},
  number  = {3},
  pages   = {267--283}
}

@book{Bronshtein2015handbook,
  author    = {Bronshtein, Ilja N. and Semendyayev, Konstantin A. and Musiol, Gerhard and M\"{u}hlig, Heiner},
  title     = {Handbook of Mathematics},
  edition   = {6th},
  publisher = {Springer},
  year      = {2015},
  address   = {Berlin, Heidelberg},
  isbn      = {978-3-662-46220-1}
}

@book{Ince1956ordinary,
  title={Ordinary Differential Equations},
  author={Ince, Edward Lindsay},
  year={1956},
  publisher={Dover Publications},
  address={New York}
}

@book{Schinzinger1991,
  author    = {Schinzinger, Roland and Laura, Patricio A. A.},
  title     = {Conformal Mapping: Methods and Applications},
  publisher = {Elsevier Science Publishers},
  year      = {1991},
  address   = {Amsterdam},
  isbn      = {9780444888068}
}

@article{Konoplya:2022zav,
    author = "Konoplya, R. A. and Zhidenko, A.",
    title = "{Bernstein spectral method for quasinormal modes of a generic black hole spacetime and application to instability of dilaton{\textendash}de Sitter solution}",
    eprint = "2211.02997",
    archivePrefix = "arXiv",
    primaryClass = "gr-qc",
    doi = "10.1103/PhysRevD.107.044009",
    journal = "Phys. Rev. D",
    volume = "107",
    number = "4",
    pages = "044009",
    year = "2023"
}

@article{Konoplya2011RMP,
    author = "Konoplya, R. A. and Zhidenko, A.",
    title = "{Quasinormal modes of black holes: From astrophysics to string theory}",
    eprint = "1102.4014",
    archivePrefix = "arXiv",
    primaryClass = "gr-qc",
    doi = "10.1103/RevModPhys.83.793",
    journal = "Rev. Mod. Phys.",
    volume = "83",
    pages = "793--836",
    year = "2011"
}

@article{Lagos2020_Anomalous,
    title = {Anomalous decay rate of quasinormal modes},
    volume = {101},
    journal = {Phys. Rev. D},
    doi = {10.1103/PhysRevD.101.084018},
    number = {8},
    author = {Lagos, Macarena and Ferreira, Pedro G. and Tattersall, Oliver J.},
    year = {2020},
    note = {arXiv: 2002.01897 [gr-qc]},
    pages = {084018},
}

@article{Jansen:2017_Overdamped,
    title = {Overdamped modes in {Schwarzschild}-de {Sitter} and a {Mathematica} package for the numerical computation of quasinormal modes},
    volume = {132},
    doi = {10.1140/epjp/i2017-11825-9},
    number = {12},
    journal = {Eur. Phys. J. Plus},
    author = {Jansen, Aron},
    year = {2017},
    note = {arXiv: 1709.09178 [gr-qc]},
    pages = {546},
}

@article{DiasSantos2020_RNdSinstability,
    title = {Origin of the reissner-nordström–de sitter instability},
    journal = {Phys. Rev. D},
    volume = {102},
    doi = {10.1103/PhysRevD.102.124039},
    number = {12},
    author = {Dias, Oscar J. C. and Santos, Jorge E.},
    year = {2020},
    note = {arXiv: 2005.03673 [hep-th]},
    pages = {124039},
}

@book{BritoCardosoPani2020_Superradiance,
    title = {Superradiance},
    url = {https://doi.org/10.1007%2F978-3-030-46622-0},
    doi = {10.1007/978-3-030-46622-0},
    publisher = {Springer International Publishing},
    author = {Brito, Richard and Cardoso, Vitor and Pani, Paolo},
    year = {2020},
    note = {arXiv: 1501.06570 [gr-qc]},
}

@article{Hoof2026MNRAS,
  author        = {Hoof, Sebastian and Marsh, David J. E. and Sisk-Reyn{\'e}s, J{\'u}lia and Matthews, James H. and Reynolds, Christopher},
  title         = {Getting More Out of Black Hole Superradiance: A Statistically Rigorous Approach to Ultralight Boson Constraints from Black Hole Spin Measurements},
  journal       = {Mon. Not. R. Astron. Soc.},
  volume        = {546},
  pages         = {staf1564},
  year          = {2026},
  eprint        = {2406.10337},
  archivePrefix = {arXiv},
  primaryClass  = {hep-ph},
  doi           = {10.1093/mnras/staf1564}
}

@article{LVK2022ScalarCloud,
  author        = {{LIGO Scientific Collaboration} and {Virgo Collaboration} and {KAGRA Collaboration}},
  title         = {All-Sky Search for Gravitational Wave Emission from Scalar Boson Clouds around Spinning Black Holes in {LIGO} {O3} Data},
  journal       = {Phys. Rev. D},
  volume        = {105},
  pages         = {102001},
  year          = {2022},
  eprint        = {2111.15507},
  archivePrefix = {arXiv},
  primaryClass  = {astro-ph.HE},
  doi           = {10.1103/PhysRevD.105.102001}
}

@article{WitteMummery2025PRD,
  author        = {Witte, Samuel J. and Mummery, Andrew},
  title         = {Stepping Up Superradiance Constraints on Axions},
  journal       = {Phys. Rev. D},
  volume        = {111},
  number        = {8},
  pages         = {083044},
  year          = {2025},
  eprint        = {2412.03655},
  archivePrefix = {arXiv},
  primaryClass  = {hep-ph},
  doi           = {10.1103/PhysRevD.111.083044}
}

@misc{NingSafdiWelch2026GWTC5,
  author        = {Ning, Orion and Safdi, Benjamin R. and Welch, Catherine},
  title         = {No Evidence for Superradiant Axions in {LIGO}--{Virgo}--{KAGRA} {GWTC-5} Binary Black Hole Spins},
  year          = {2026},
  eprint        = {2607.01317},
  archivePrefix = {arXiv},
  primaryClass  = {hep-ph},
  note          = {arXiv:2607.01317 [hep-ph]}
}

@article{ZuFengYuanFan2020EPJP,
  author        = {Zu, Lei and Feng, Lei and Yuan, Qiang and Fan, Yi-Zhong},
  title         = {Stringent Constraints on the Light Boson Model with Supermassive Black Hole Spin Measurements},
  journal       = {Eur. Phys. J. Plus},
  volume        = {135},
  pages         = {709},
  year          = {2020},
  eprint        = {2007.03222},
  archivePrefix = {arXiv},
  primaryClass  = {astro-ph.HE},
  doi           = {10.1140/epjp/s13360-020-00734-9}
}

@article{BaticDutykhSukaiti2026KSQC,
  author        = {Batic, Davide and Dutykh, Denys and Sukaiti, Mark Essa},
  title         = {Quasinormal modes of the Kazakov--Solodukhin quantum-corrected black hole: a spectral analysis},
  journal       = {Eur. Phys. J. C},
  volume        = {86},
  pages         = {847},
  year          = {2026},
  doi           = {10.1140/epjc/s10052-026-16102-3}
}

@article{KoyamaTomimatsu2001MassiveTail,
  author        = {Koyama, Hiroko and Tomimatsu, Akira},
  title         = {Asymptotic Tails of Massive Scalar Fields in Schwarzschild Background},
  journal       = {Phys. Rev. D},
  volume        = {64},
  pages         = {044014},
  year          = {2001},
  eprint        = {gr-qc/0103086},
  archivePrefix = {arXiv},
  primaryClass  = {gr-qc},
  doi           = {10.1103/PhysRevD.64.044014}
}

@article{QianEtAl2022MassiveTails,
  author        = {Qian, Wei-Liang and Lin, Kai and Shao, Cai-Ying and Wang, Bin and Yue, Rui-Hong},
  title         = {On the Late-Time Tails of Massive Perturbations in Spherically Symmetric Black Holes},
  journal       = {Eur. Phys. J. C},
  volume        = {82},
  pages         = {931},
  year          = {2022},
  eprint        = {2203.04477},
  archivePrefix = {arXiv},
  primaryClass  = {gr-qc},
  doi           = {10.1140/epjc/s10052-022-10910-z}
}

@article{AlvesPonquioMedeiros2025PRD,
  author        = {Alves, Matheus F. S. and P{\^o}nquio, Bruno P. and Medeiros, L. G.},
  title         = {Critical masses and numerical computation of massive scalar quasinormal modes in Schwarzschild black holes},
  journal       = {Phys. Rev. D},
  volume        = {112},
  pages         = {124007},
  year          = {2025},
  eprint        = {2509.07235},
  archivePrefix = {arXiv},
  primaryClass  = {gr-qc},
  doi           = {10.1103/rjrk-j3jt}
}

\end{document}